\documentclass{jfm}

\usepackage{graphicx, color}
\usepackage{natbib}
\usepackage{amssymb}
\usepackage{upmath}
\usepackage{amsbsy}
\usepackage{amsmath}
\usepackage{verbatim}
\usepackage{lscape}
\usepackage{fancyhdr}

  \usepackage[pdftex, breaklinks=true, colorlinks=true, citecolor=blue, linkcolor=blue, urlcolor=blue, bookmarksdepth=3, pdftitle={Optimal Taylor-Couette turbulence}]{hyperref}

\newcommand\aOpt{\mbox{$a_{opt}$}}  
\newcommand{\be}{\begin{equation}}
\newcommand{\ee}{\end{equation}}

\def\r{{\mbox{\boldmath$r$}}}

\title[Optimal Taylor-Couette turbulence]{Optimal Taylor-Couette turbulence}

\author[D. P. M. van Gils, S. G. Huisman, S. Grossmann, C. Sun, and D. Lohse]
{Dennis P.\ M.\ van Gils$^1$, Sander G.\ Huisman$^1$, Siegfried Grossmann$^2$, Chao Sun$^1$
and Detlef Lohse$^1$\footnote{Email address for correspondence: \href{mailto:d.lohse@utwente.nl}{d.lohse@utwente.nl}}}

\affiliation{$^1$ Physics of Fluids Group, Faculty of Science and Technology, Impact \& MESA+ Institutes, and Burgers Center for Fluid Dynamics, University of Twente, 7500AE Enschede, The Netherlands\\[\affilskip]
$^2$ Department of Physics, Renthof 6, University of Marburg, D-35032 Marburg, Germany}

\pubyear{2011}
\volume{?}
\pagerange{??}
\date{27 November 2011; revised 31 March 2012; accepted 15 May 2012}
\begin{document}

\maketitle
\thispagestyle{fancy}

\begin{abstract}
Strongly turbulent Taylor-Couette flow with independently rotating inner and outer cylinders with a radius ratio of $\eta = 0.716$ is experimentally studied. From global torque measurements, we analyse the dimensionless angular velocity flux $Nu_\omega (Ta, a)$ as a function of the Taylor number $Ta$ and the angular velocity ratio $a = - \omega_o/\omega_i$ in the large-Taylor-number regime $10^{11} \lesssim Ta \lesssim 10^{13}$ and well off the inviscid stability borders (Rayleigh lines) $a = - \eta^2$ for co-rotation and $a = \infty$ for counter-rotation. We analyse the data with the common power-law ansatz for the dimensionless angular velocity transport flux $Nu_\omega (Ta, a) = f(a) Ta^{\gamma}$, with an amplitude $f(a)$ and an exponent $\gamma$. The data are consistent with one effective exponent $\gamma = 0.39 \pm 0.03$ for all $a$, but we discuss a possible $a$ dependence in the co- and weakly counter-rotating regimes. The amplitude of the angular velocity flux $f(a) \equiv Nu_{\omega}(Ta,a) / Ta^{0.39}$ is measured to be maximal at slight counter-rotation, namely at an angular velocity ratio of $a_{opt} = 0.33 \pm 0.04$, i.e.\ along the line $\omega_o = -0.33 \omega_i$. This value is theoretically interpreted as the result of a competition between the destabilizing inner cylinder rotation and the stabilizing but shear-enhancing outer cylinder counter-rotation. With the help of laser Doppler anemometry, we provide angular velocity profiles and in particular identify the radial position $r_n$ of the neutral line, defined by $\langle \omega (r_n) \rangle_t = 0$ for fixed height $z$. For these large $Ta$ values the ratio $a \approx 0.40$, which is close to $a_{opt} = 0.33$, is distinguished by a zero angular velocity gradient $\partial \omega /\partial r = 0$ in the bulk. While for moderate counter-rotation $-0.40 \omega_i \lesssim \omega_o < 0$, the neutral line still remains close to the outer cylinder and the probability distribution function of the bulk angular velocity is observed to be monomodal. For stronger counter-rotation the neutral line is pushed inwards towards the inner cylinder; in this regime the probability distribution function of the bulk angular velocity becomes bimodal, reflecting intermittent bursts of turbulent structures beyond the neutral line into the outer flow domain, which otherwise is stabilized by the counter-rotating outer cylinder. Finally, a hypothesis is offered allowing a unifying view and consistent interpretation for all these various results.

\noindent\centering{doi: \href{http://dx.doi.org/10.1017/jfm.2012.236}{10.1017/jfm.2012.236}}
\end{abstract}

\bigskip
\hrule

\section{Introduction}\label{sec1}

Taylor-Couette (TC) flow (the flow between two coaxial, independently rotating cylinders) is, next to Rayleigh-B\'enard (RB) flow (the flow in a box heated from below and cooled from above), the most prominent `Drosophila' on which to test hydrodynamic concepts for flows in closed containers. For outer cylinder rotation and fixed inner cylinder, the flow is linearly stable. In contrast, for inner cylinder rotation and fixed outer cylinder, the flow is linearly unstable thanks to the driving centrifugal forces (see e.g.\ \citet{tay23,col65,pfi81,pri81,smi82,mul82,mul87,pfi88,buc96,ess96}). The case of two independently rotating cylinders has been well analysed for low Reynolds numbers (see e.g.\ \cite{and86}). For large Reynolds numbers, where the bulk flow is turbulent, studies have been scarce -- see, for example, the historical work by \cite{wen33} or the experiments by \citet{and86, richard2001, dub05, bor10, rav10, hou11}. \cite{ji06,bur10} examined the local angular velocity flux with laser Doppler anemometry in independently rotating cylinders at high Reynolds numbers (to be defined below) up to $2\times 10^6$. Recently, in two independent experiments, \cite{gil11} and \cite{pao11} supplied precise data for the global torque scaling in the turbulent regime of the flow between independently rotating cylinders.

We use cylindrical coordinates $r, \phi$ and $z$. Next to the geometric ratio $\eta = r_i /r_o$ between the inner cylinder radius $r_i$ and the outer cylinder radius $r_o$, and the aspect ratio $\Gamma = L / d$ of the cell height $L$ and the gap width $d= r_o- r_i$, the dimensionless control parameters of the system can be expressed either in terms of the inner and outer cylinder Reynolds numbers $Re_i = r_i \omega_i d /\nu $ and
$Re_o = r_o \omega_o d /\nu $, respectively, or in terms of the ratio of the angular velocities
\be
a = - \omega_o/\omega_i
\ee
and the Taylor number
\be
Ta = \frac{1}{4} \sigma (r_o - r_i)^2 (r_i+r_o)^2 (\omega_i - \omega_o)^2  \nu^{-2}.
\label{eq:Taylor}
\ee
Here, according to the theory by \citet*{eck07b} (from now on called EGL), $\sigma = (((1+\eta)/2)/\sqrt{\eta})^4$ (thus $\sigma = 1.057$ for the current $\eta = 0.716$ of the used TC facility) can be formally interpreted as a `geometrical' Prandtl number and $\nu$ is the kinematic viscosity of the fluid. With $r_a = (r_i + r_o)/2$ and $r_g = \sqrt{r_ir_o}$, the arithmetic and the geometric mean radii, the Taylor number can be written as
\be
Ta = r_a^6 r_g^{-4} d^2 \nu^{-2} (\omega_i - \omega_o)^2.
\label{eq:Taylor2}
\ee
The angular velocity of the inner cylinder $\omega_i$ is always defined as positive, whereas the angular velocity of the outer cylinder $\omega_o$ can be either positive (co-rotation) or negative (counter-rotation). Positive $a$ thus refers to the counter-rotating case on which our main focus will lie.

The response of the system is the degree of turbulence of the flow between the cylinders (e.g.\ expressed in a wind Reynolds number of the flow, measuring the amplitude of the $r$ and $z$ components of the velocity field) and the torque $\tau$ that is necessary to keep the inner cylinder rotating at constant angular velocity. Following the suggestion of EGL, the torque can be non-dimensionalized in terms of the laminar torque to define the (dimensionless) `Nusselt number'
\be
Nu_\omega = \frac{\tau}{2\pi L \rho_{fluid} J_{lam}^\omega},
\label{nu_omega}
\ee
where $\rho_{fluid}$ is the density of the fluid between the cylinders and
 \be
J^\omega_{lam} = 2\nu r_i^2 r_o^2 { \omega_i - \omega_o \over  r_o^2 - r_i^2}
\label{jlam}
\ee
is the conserved angular velocity flux in the laminar case. The reason for the choice (\ref{nu_omega}) is that
\be
J^\omega = J^\omega_{lam} Nu_\omega =
 r^3 \left( \left< u_r \omega \right>_{A,t} - \nu \partial_r \left< \omega \right>_{A,t} \right)
\label{eq:J_omega}
\ee
is the relevant conserved transport quantity, representing the flux of angular velocity from the inner to the outer cylinder. This definition of $J^{\omega}$ is an immediate consequence of the Navier-Stokes equations. (The authors would like to point out that (\ref{eq:J_omega}) appeared first, in a different notation, in \cite{bus72}, where $J^{\omega}$ was called the `torque'. Equations (3.4) and (4.13) of \cite{eck07b} are analogous to (3.2) and (3.4) of \cite{bus72}.) Here $u_r$ ($u_\phi$) is the radial (azimuthal) velocity, $\omega = u_\phi/r$ the angular velocity, and $\left< \dots\right>_{A,t}$ characterizes averaging over time and a cylindrical surface with constant radius $r$. With this choice of control and response parameters, EGL could work out a close analogy between turbulent TC and turbulent RB flow, building on \citet{gro00} and extending the earlier work of \cite{bra69} and \cite{dub02}. This was further elaborated by \cite{gil11}.

The main findings of \cite{gil11}, who operated the TC set-up, known as the Twente turbulent Taylor-Couette system or T$^3$C, at fixed $\eta = 0.716$ and for $Ta > 10^{11}$ as well as the variable $a$ well off the stability borders $-\eta^2$ and $\infty$, are as follows: (i) in the $(Ta, a)$ representation, $Nu_\omega (Ta, a)$ within the experimental precision factorizes into $Nu_\omega (Ta, a) = f(a) F(Ta)$; (ii) $F(Ta) = Ta^{0.38}$ for all analysed $-0.4 \le  a \le 2.0$ in the turbulent regime; and (iii) $f(a) = Nu_\omega (Ta, a)/ Ta^{0.38}$ has a pronounced maximum around $\aOpt \approx 0.4$. Also \cite{pao11}, at slightly different $\eta = 0.725$, found such a maximum in $f(a)$, namely at $\aOpt \approx 0.35$. For this $\aOpt$ the angular velocity transfer amplitude $f(\aOpt(\eta))$ for the transport from the inner to the outer cylinder is maximal. -- From these findings one has to conclude that, for not too strong counter-rotation $-0.4\omega_i \lesssim \omega_o < 0$, the angular velocity transport flux is still further enhanced as compared to the case of fixed outer cylinder $\omega_o=0$. This stronger turbulence is attributed to the enhanced shear between the counter-rotating cylinders. Only for strong enough counter-rotation $\omega_o < -\aOpt \omega_i$ (i.e.\ $a > \aOpt$) does the stabilization through the counter-rotating outer cylinder take over and the transport amplitude decrease with further increasing $a$.

The aims of this paper are to provide further and more precise data on the maximum in the conserved turbulent angular velocity flux $Nu_\omega (Ta, a) / Ta^{\gamma} = f(a)$ as a function of $a$ and a theoretical interpretation of this maximum, including a speculation on how it depends on $\eta$. We also put our findings in the perspective of the earlier results on highly turbulent TC flow by \cite{lat92,lat92a} and \cite{lew99} and on recent results on highly turbulent Rayleigh-B\'enard flow by \cite{he12}: We think that all these experiments achieve the so-called ``ultimate regime'' in which the boundary layers are already turbulent. Next we provide laser Doppler anemometry (LDA) measurements of the angular velocity profiles $\langle \omega(r) \rangle_t$ as functions of height, and show that the flow close to the maximum in $f(a)$, for these asymptotic $Ta$ and deep in the instability range at $a = \aOpt$, has a vanishing angular velocity gradient $\partial \omega / \partial r$ in the bulk of the flow. We identify the location of the neutral line $r_n$, defined by $\langle \omega(r_n) \rangle_t = 0$ for fixed height $z$, finding that it remains still close to the outer cylinder $r_o$ for weak counter-rotation, $0< a < \aOpt$, but starts moving inwards towards the inner cylinder $r_i$ for $a  \gtrsim \aOpt$. Finally we show that the turbulent flow organization totally changes for {$a \gtrsim \aOpt$}, where the stabilizing effect of the strong counter-rotation reduces the angular velocity transport. In this strongly counter-rotating regime the probability distribution function of the angular velocity in the bulk becomes bimodal, reflecting intermittent bursts of turbulent structures beyond the neutral line towards the outer flow region, which otherwise, i.e.\ in between such bursts, is stabilized by the counter-rotating outer cylinder.

The outline of the paper is as follows. The experimental set-up is introduced in section \ref{sec2} and we discuss, additionally, the height dependence of the flow profile and finite size effects. The global torque results are reported and discussed  in section \ref{sec3}. Section \ref{sec4} and \ref{sec5} provide laser Doppler anemometry (LDA) measurements on the angular velocity radial profiles and on the turbulent flow structures inside the TC gap. The paper ends, in section \ref{sec7}, with a summary, further discussions of the neutral line inside the flow, and an outlook.

\begin{figure}
\centerline{\includegraphics[height=7cm]{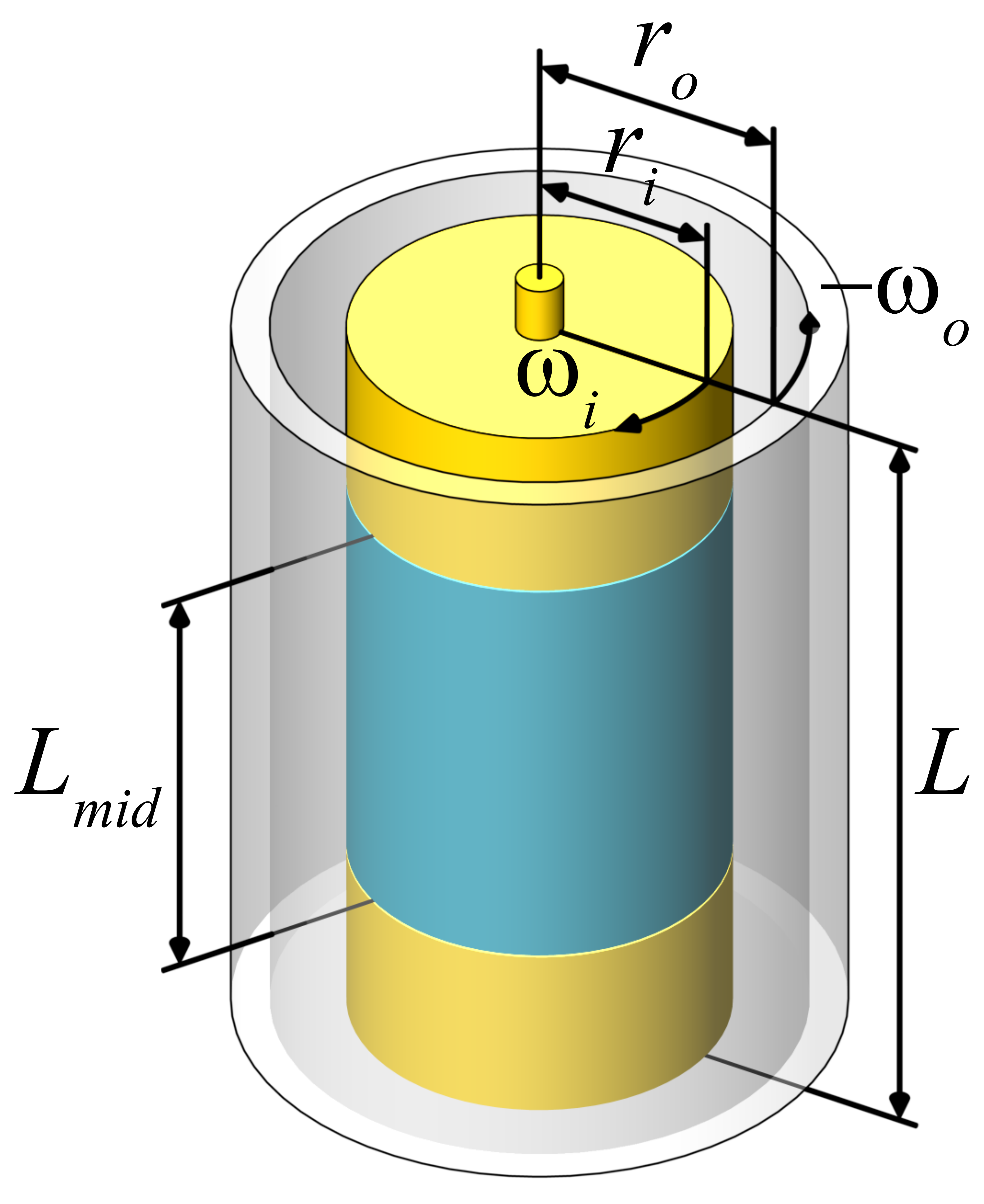}}
\caption{Sketch of the T$^3$C Taylor-Couette cell employed for the measurements presented here (from \cite{gil11a}). The total height of the cell is $L = 92.7$ cm. The torque measurements are made with the middle part of the cell with length $L_{mid} = 53.6$ cm, in order to minimize edge effects. The outer and inner cylinder radii are $r_o = 27.94$ cm and $r_i = 20.00$ cm, leading to a radius ratio of $\eta = 0.716$, a gap width of $d=r_o-r_i= 7.94$ cm, an aspect ratio of $\Gamma = L/d =  11.68$, and an internal fluid volume of $111$ litres. The inner and outer cylinder angular velocities are denoted by $\omega_i$ and $\omega_o$, respectively. By definition, $\omega_i>0$, implying that $\omega_o < 0$ or $a = - \omega_o/\omega_i >0$ represents counter-rotation, on which we focus in this paper. The top and bottom plates are attached to the outer cylinder.
\label{fig:01}}
\end{figure}

\section{Experimental setup and discussion of end-effects} \label{sec2}
The core of our experimental set-up, the Taylor-Couette cell, is shown in figure \ref{fig:01}. In the caption we give the respective length scales and their ratios. In particular, the fixed geometric dimensionless numbers are $\eta = 0.716$ and $\Gamma = 11.68$. The details of the set-up are given in \cite{gil11a}. The working liquid is water at a continuously  controlled constant temperature (precision $\pm 0.5$K) in the range $19^{\circ}$C - $26^{\circ}$C. The accuracy in setting and maintaining a constant $a$ is $\pm 0.001$ based on direct angular velocity measurements of the T$^3$C facility. To reduce edge effects, similarly as in \cite{lat92,lat92a}, the torque is measured at the middle part (length ratio $L_{mid} /L = 0.578$)  of the inner cylinder. \cite{lat92}'s original motivation for this choice was that the height of the remaining upper and lower parts of the cylinder roughly equals the size of a pair of Taylor vortices. While the respective first or last Taylor vortex indeed will be affected by the upper and lower plates (which in our T$^3$C cell are attached to the outer cylinder), the hope is that in the strongly turbulent regime the turbulent bulk is not affected by such edge effects. Note that for the laminar case (e.g.\ for pure outer cylinder rotation), this clearly is not the case, as has been known since  \cite{tay23} -- see, for example, the classical experiments by \cite{col66}, the numerical work by \cite{hol04}, or the review by \cite{tag94}. For such weakly rotating systems, profile distortions from the plates propagate into the fluid and dominate the whole laminar velocity field. The velocity profile will then be very different from the classical height-independent laminar profile (see e.g.\ \cite{ll87}) with periodic boundary conditions in the vertical direction,
\be
u_{\phi, \mathrm{lam}} = A r  + B/r  \ , \qquad A = \frac{\omega_o - \eta^2
 \omega_i}{1-\eta^2} \ , \qquad B = \frac{(\omega_i-\omega_o)r_i^2}{1 - \eta^2} \ .
\label{eq:laminar_profile}
\ee

To control edge effects and ensure that they are indeed negligible in the strongly turbulent case under consideration here ($10^{11} < Ta < 10^{13}$ and $-0.40 \le a \le 2.0$, so well off the instability borders), we have measured time series of the angular velocity $\omega ( \r , t) = u_\phi (\r ,t) / r $ for various heights $0.32 < z/L < 1$ and radial positions $r_i < r < r_o$ with laser  Doppler anemometry (LDA). We employ a backscatter LDA configuration set-up with a measurement volume of 0.07 mm x 0.07 mm x 0.3 mm. The seeding particles (PSP-5, Dantec Dynamics) have a mean radius of $r_{seed} = 2.5~\mu$m and a density of $\rho_{seed} = 1.03$ g cm$^{-3}$.

We estimate the minimum velocity difference $\Delta v = v_{seed} - v_{fluid}$ between a particle $v_{seed}$ and its surrounding fluid $v_{fluid}$ needed for the drag force $F_{drag} = 6 \pi \mu r_{seed} \Delta v$ to outweigh the centrifugal force $F_{cent}(r)=4 \pi {r_{seed}}^3 \left( \rho_{seed} - \rho_{fluid}\right) v^2 / (3r)$. We put in $v = 5$ m s$^{-1}$ as a typical azimuthal velocity inside the TC gap at mid-gap radial position $r = 0.24$ m, with $\rho_{fluid} = 1.00$ g cm$^{-3}$ as the density and $\mu = 9.8\times10^{-4}$ kg m$^{-1}$ s$^{-1}$ as the dynamic viscosity of water at 21$^{\circ}$C. This results in $\Delta v = 2{r_{seed}}^2 \left( \rho_{seed} - \rho_{fluid} \right)v^2/(9\mu r) \approx 5\times 10^{-6}$ m s$^{-1}$, which is several orders of magnitude smaller than the typical velocity fluctuation inside the TC-gap of order $10^{-1}$ m s$^{-1}$, and hence centrifugal forces on the seeding particles are negligible.

\begin{figure}
\centerline{\includegraphics[width=.9\textwidth]{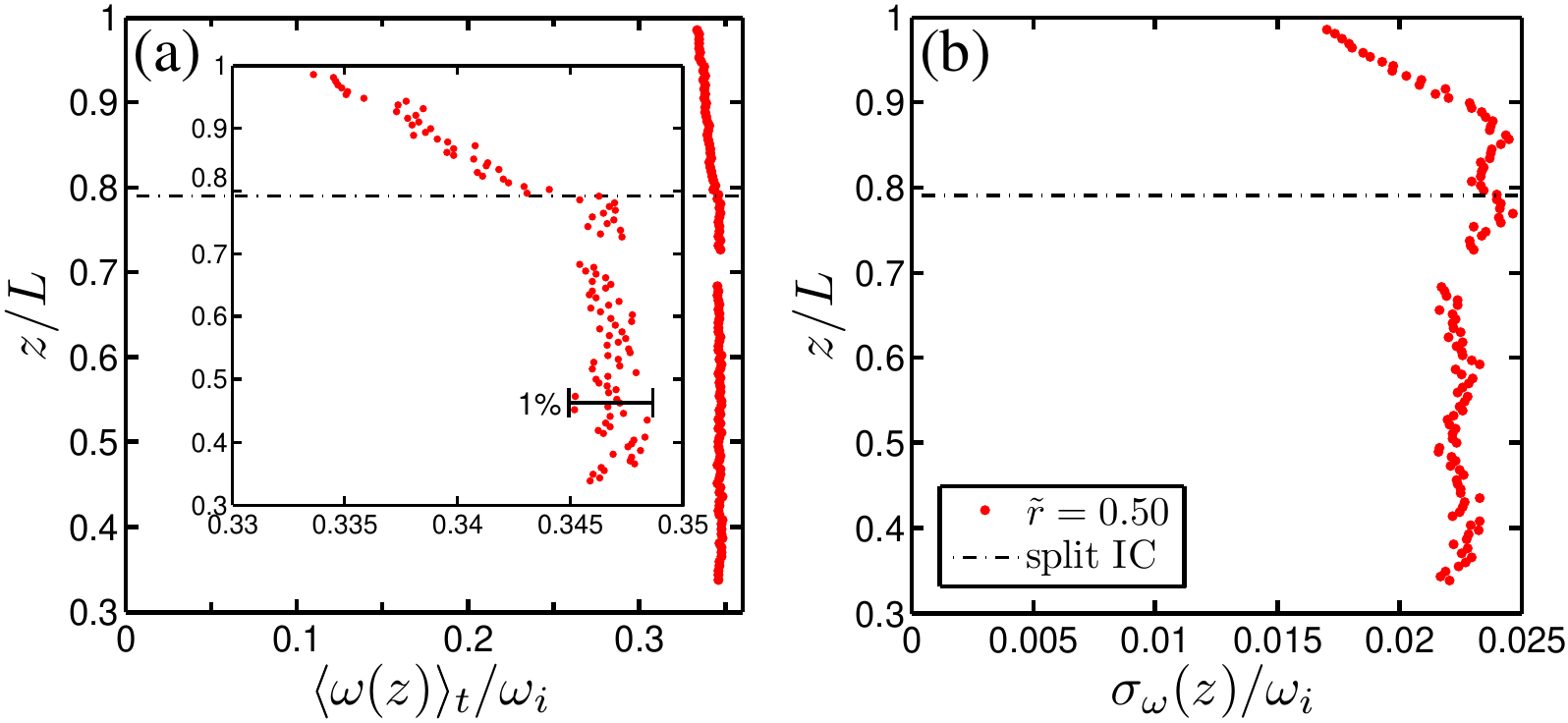}}
\caption{Time-averaged axial profiles of the azimuthal angular velocity inside the T$^3$C measured with LDA at mid-gap, i.e.\ $\tilde{r}=(r - r_i)/(r_o - r_i)=0.5$, for the case $Re_i=1.0\times10^6$ and $Re_o=0$ (corresponding to $a=0$ and $Ta = 1.5 \times 10^{12}$). The height $z$ from the bottom plate is normalized against the total height $L$ of the inner volume of the tank. (\emph{a}) The time-averaged angular velocity $\langle\omega(z, \tilde{r}=1/2)\rangle_t$ normalized by the angular velocity of the inner cylinder wall $\omega_i$. (\emph{b}) The standard deviation of the angular velocity $\sigma_{\omega}(z)$ normalized by the angular velocity of the inner wall. The split between the middle and the top inner cylinder sections is indicated by the dashed-dotted line at $z/L=0.79$. As can be appreciated in this figure, the end-effects are negligible over the middle section where we measure the global torques as reported in this work. The velocities near the top plate, $z/L = 1$, are not sufficiently resolved to see the boundary layer.
\label{fig:02}}
\end{figure}

We account for the refraction due to the cylindrical interfaces -- details are given by \citet{hui12b}. Figure \ref{fig:02}(\emph{a}) shows the height dependence of the time-averaged angular velocity at mid-gap, $\tilde{r}=(r - r_i)/(r_o - r_i) = 1/2$, for $a=0$ and $Ta = 1.5 \times 10^{12}$, corresponding to $Re_i = 1.0 \times 10^6$ and $Re_o = 0$. The dashed-dotted line at $z/L=0.79$ corresponds to the gap between the middle part of the inner cylinder, with which we measure the torque, and the upper part. Along the middle part the time-averaged angular velocity is $z$ independent within 1\%, as is demonstrated in the inset, showing the enlarged relevant section of the $\omega$ axis. From the upper edge of the middle part of the inner cylinder towards the highest position that we can resolve, $0.5$ mm below the top plate, the mean angular velocity decays by only 5\%. This finite difference might be due to the existence of Ekman layers near the top and bottom plate (\citet{gre90}). Since at $z/L=1$ we have $\omega (r,t) = 0$, as the upper plate is at rest for $a=0$ or $\omega_o=0$, 95\% of the edge effects on $\omega$ occur in such a thin fluid layer near the top (bottom) plate that we cannot even resolve it with our present LDA measurements.

\begin{figure}
\centerline{\includegraphics[width=.9\textwidth]{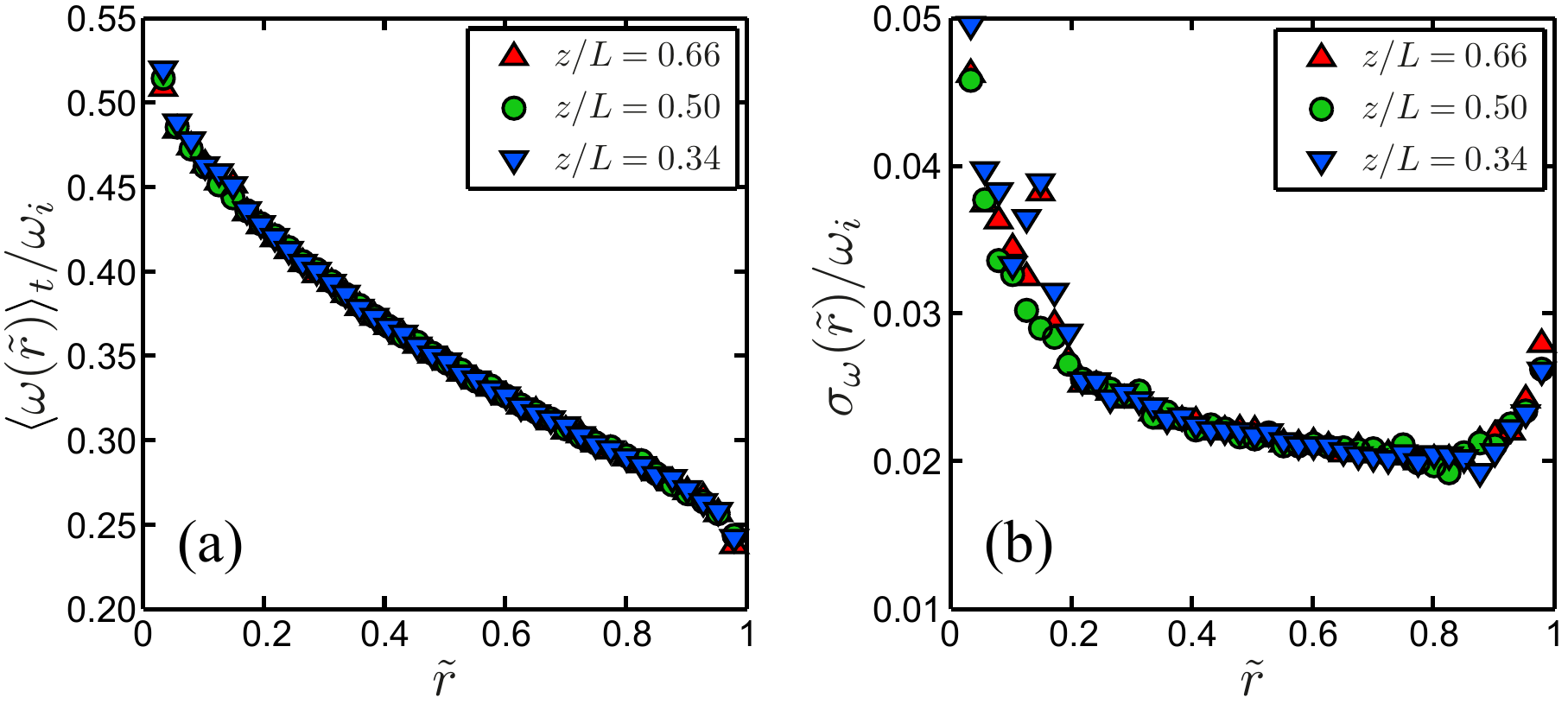}}
\caption{Radial profiles of the azimuthal angular velocity as presented in figure \ref{fig:02}, scanned at three different heights $z/L$ = 0.66, 0.50 and 0.34, plotted against the dimensionless gap distance $\tilde{r}=(r - r_i)/(r_o - r_i)$, again for $Re_o = 0$ and $Re_i = 1 \times 10^6$. (\emph{a}) The time-averaged angular velocity $\langle\omega(\tilde{r})\rangle_t$ normalized by the angular velocity of the inner wall $\omega_i$. All profiles fall on top of each other, showing no axial dependence of the flow in the investigated axial range. (\emph{b}) Standard deviation of the angular velocity $\sigma_{\omega}(\tilde{r})$ normalized by the angular velocity of the inner wall. The velocity fluctuations show no significant axial dependence in  the investigated axial range. The boundary layers at the cylinder walls are not resolved.
\label{fig:03}}
\end{figure}

For the angular velocity fluctuations shown in figure \ref{fig:02}(\emph{b}), we observe a 25\% decay in the upper 10\% of the cylinder, but again in the measurement section of the inner cylinder $0.2< z/L < 0.8$ there are no indications of any edge effects. The plots of figure \ref{fig:02} together confirm that edge effects are unlikely to  play a visible role for our torque measurements in the middle part of the cylinder. Even the Taylor vortex roll structure, which dominates TC flow at low Reynolds numbers (\cite{dom84,dom86,and86,tag94}), is not visible at all in the time-averaged angular velocity profile $\langle \omega (z, \tilde r=1/2, t)\rangle_t$.

To double check that this $z$ independence holds not only at mid-gap $\tilde{r} = 1/2$ but also for the whole radial $\omega$ profiles, we measured time series of $ \omega (z, \tilde r , t) $ at three different heights $z/L= 0.34$, $0.50$ and $0.66$. The radial dependence of the mean value and of the fluctuations are shown in figure \ref{fig:03}. The profiles are basically identical for the three heights, with the only exception of some small irregularity in the fluctuations at $z/L= 0.34$ in the small region $0.1 < \tilde{r} < 0.2$, whose origin is unclear to us. Note that in both panels of figure \ref{fig:03} the radial inner and outer boundary layers are again not resolved; in this paper we will focus on bulk properties and global scaling relations.

Based on the results of this section, we feel confident to claim that: (i) edge effects are unimportant  for the global torque measurements done with the middle part of the inner cylinder reported in section \ref{sec3}; and (ii) the local profile and fluctuation measurements done
close to mid-height $z/L = 0.44$, which will be shown and analysed in sections \ref{sec4} and \ref{sec5}, are representative for any height in the middle part of the cylinder.

\section{Global torque measurements} \label{sec3}

\begin{figure}
\centerline{\includegraphics[width=1\textwidth]{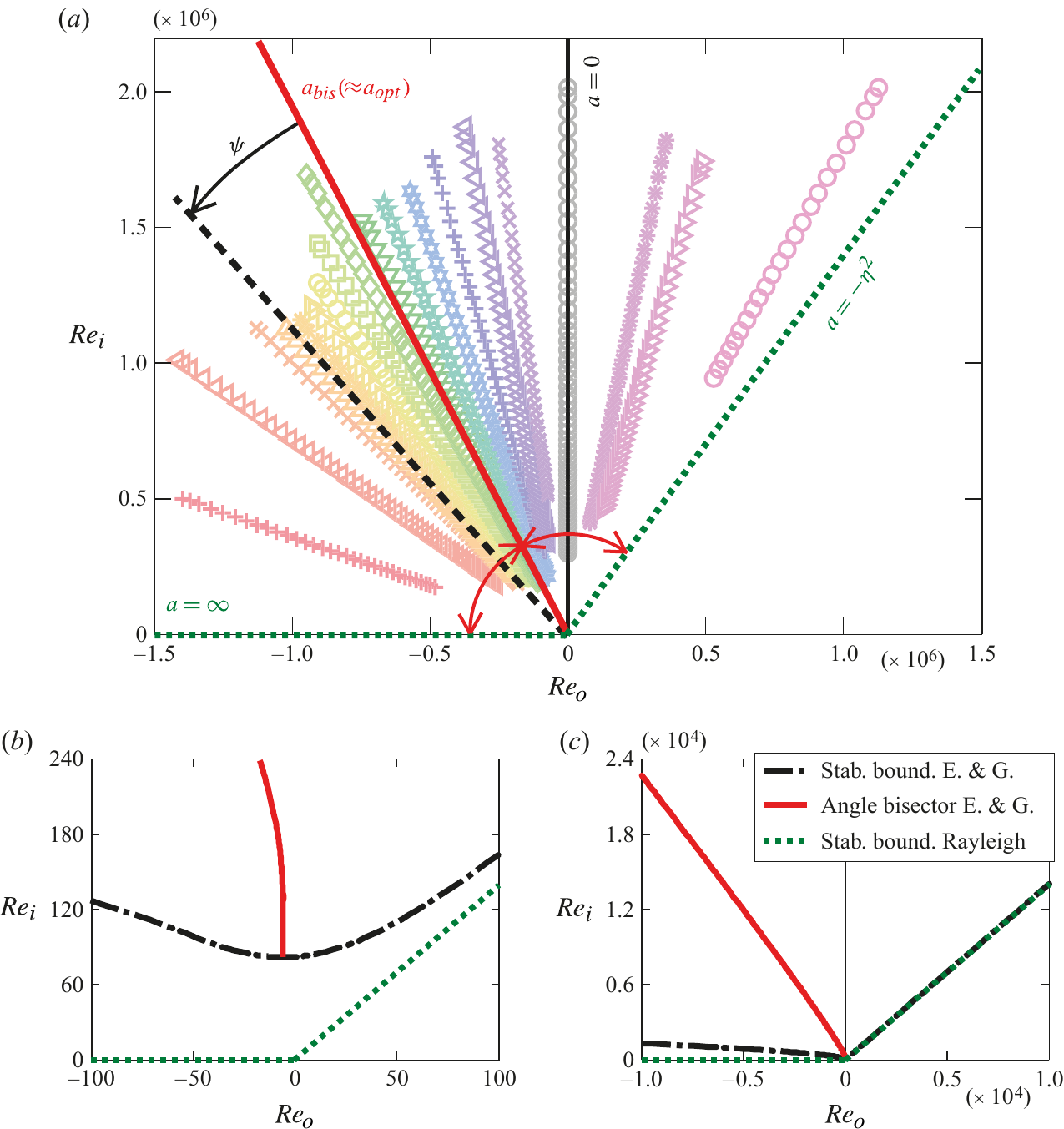}}
\caption{(a) Reynolds number phase space showing the explored regime of the T$^3$C as symbols with pale colours. The dotted green lines are the boundaries between the unstable (upper left) and stable (lower right) flow region, shown here for the radius ratio $\eta = 0.716$ as experimentally examined in this work. The green line in the right quadrant is the analytical expression for the stability boundary as found by \cite{ess96}, which recovers to the Rayleigh stability criterion $Re_o/Re_i = \eta$ for $Re_i, Re_o \gg 1$, the viscous corrections decreasing $\propto Re_o^{-2}$. The green line in the left quadrant also follows the stability boundary by Esser and Grossmann ($Re_i \propto Re_o^{3/5}$), but is taken here as $Re_o = 0$. This inviscid approximation is sufficient, if $a$ is not too large, i.e.\ away from the stability curve. Similar to \cite{gil11}, we define the parameter $a\equiv-\omega_o/\omega_i$ as the (negative) ratio between the angular rotation rates of the outer and inner cylinders. We hypothesize maximum instability and hence optimal turbulence on the bisector of the unstable region, indicated by the solid red line. (b, c) Enlargements of the $Re$ space at different scales showing the curvatures of the stability boundaries and the corresponding bisector (red). Above $Re_i, Re_o > 10^5$ the viscous deviation from straight lines becomes negligible.
\label{fig:04}}
\end{figure}

\begin{figure}
\centerline{\includegraphics[width=1\textwidth]{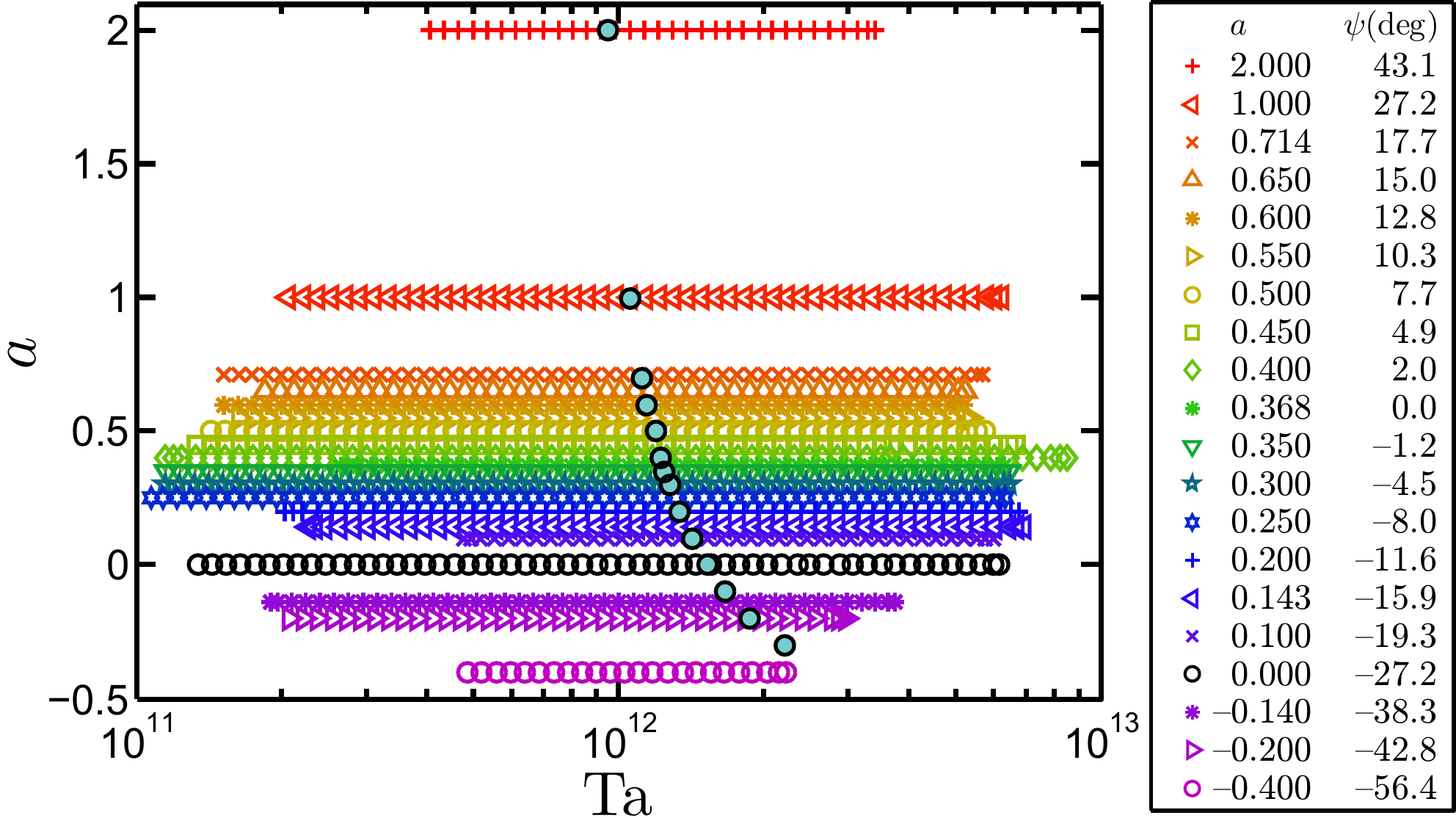}}
\caption{The probed ($Ta,a$) parameter space, equivalent to the ($Re_i,Re_o$) space shown in figure \ref{fig:04}. Each horizontal data line corresponds to a global torque measurement on the middle section of the inner cylinder at different constant $a$ (and hence $\psi$). The (blue) filled  circles correspond to local measurements on the angular velocity at fixed $Ta$ and $a$ as will be discussed in section \ref{sec4}.
\label{fig:05}}
\end{figure}

In this section we will present our data from the global torque measurements for independent inner and outer cylinder rotation, which complement and improve the precision of our earlier measurements in \cite{gil11}. The data as functions of the respective pairs of control parameters ($Ta, a$) or ($Re_i$, $Re_o$) for which we performed our measurements are given in tabular form in table \ref{tab:G-data} and in graphical form in figures \ref{fig:04}(\emph{a}) and \ref{fig:05}.

\begin{figure}
\centerline{\includegraphics[width=0.8\textwidth]{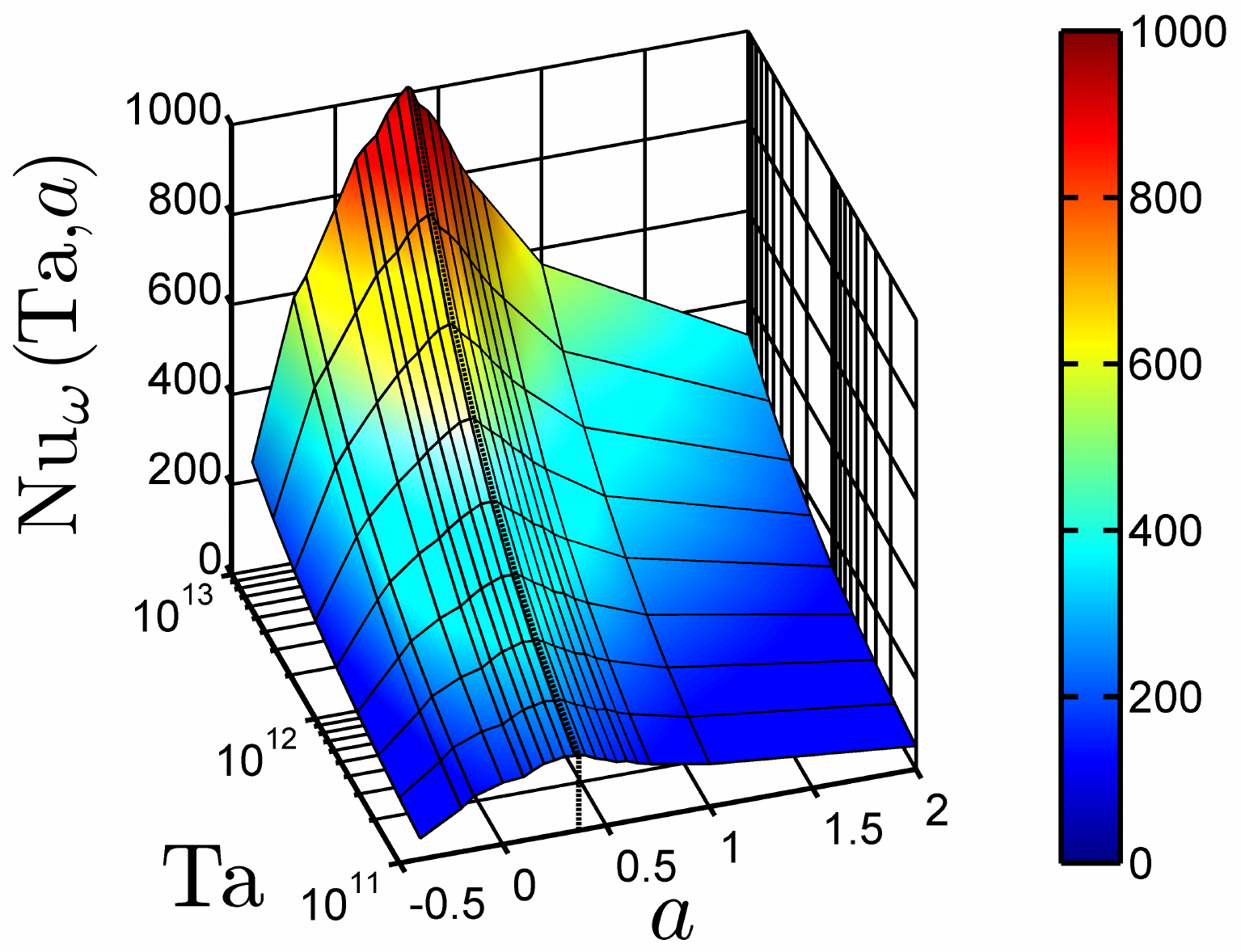}}
\caption{Three-dimensional (interpolated and extrapolated) overview
$Nu_\omega(Ta,a)$ of our experimental results. The colour and the height correspond to the $Nu_\omega$ value.
\label{fig:06}}
\end{figure}

\begin{figure}
\centerline{\includegraphics[width=1.0\textwidth]{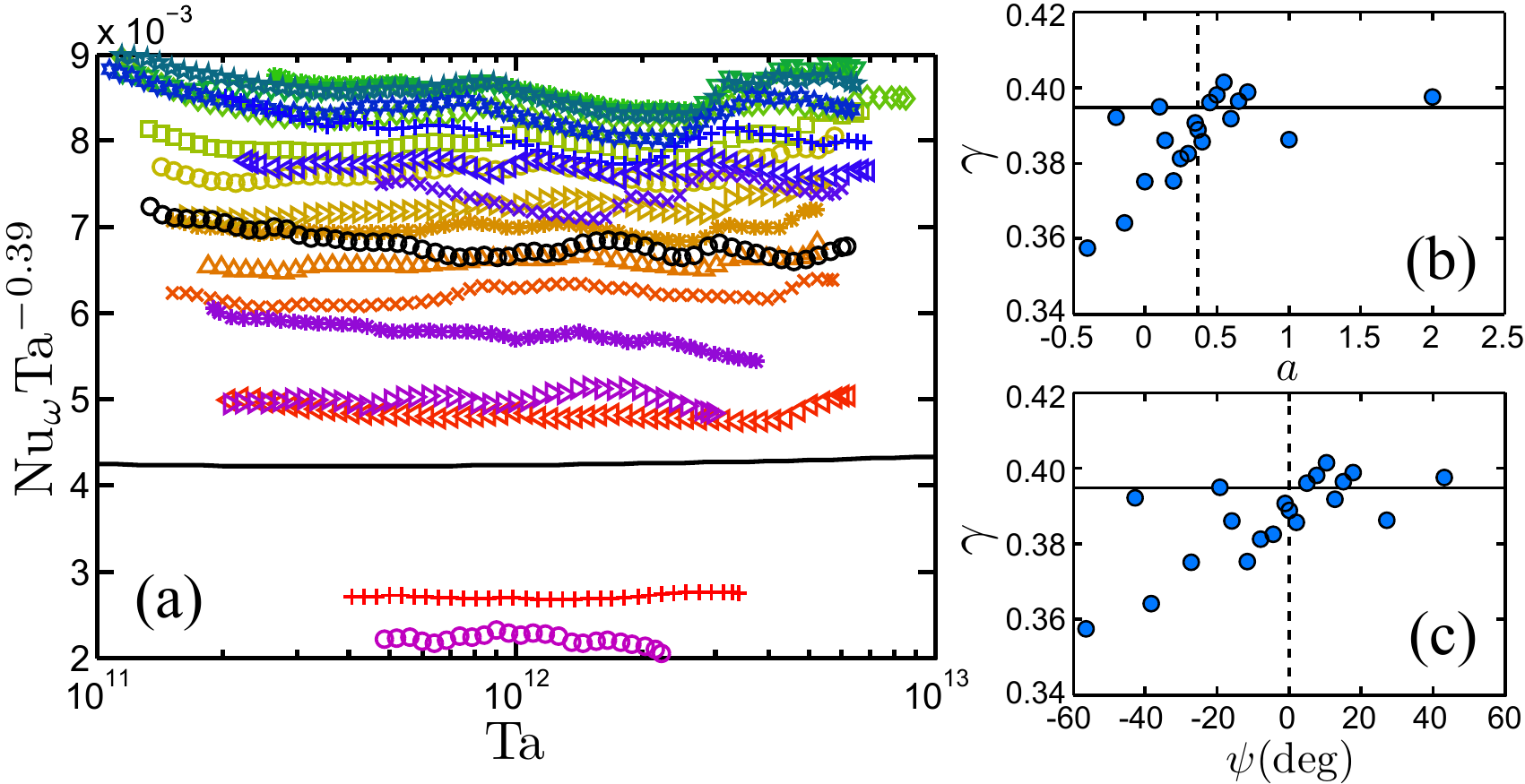}}
\caption{(\emph{a}) Plot of $Nu_\omega (Ta ,a)$, compensated by $Ta^{0.39}$, for various $a$ as a function of $Ta$, revealing effective universal scaling. The coloured symbols follow the same coding as given in the legend of figure \ref{fig:05}. The solid line has the predicted exponent from (\ref{nu-log-cor}) (cf. \cite{gro11}) and can be arbitrarily shifted in the vertical direction. Here we have used $Re_w = 0.0424 \times Ta ^{0.495}$ as found by \cite{hui12} for the case $a = 0$ over the range $4\times 10^9 < Ta < 6\times 10^{12}$, and the von K\'arm\'an constant $\bar{\kappa} = 0.4$ and $b = 0.4$, resulting in a predicted exponent of 0.395 (solid line in panels (\emph{b}) and (\emph{c})). (\emph{b, c}) The $Nu_{\omega}(Ta)$ exponent for each of the individual line series, fitted by a least-squares linear fit in log-log space, is plotted (\emph{b}) versus $a$ and (\emph{c}) versus $\psi$. Assuming $a$ independence, the average scaling exponent is $\gamma = 0.39 \pm 0.03$, which is very consistent with the effective exponent $\gamma = 0.387$ of the first-order fit on log$_{10}(Nu_{\omega})$ versus log$_{10}(Ta)$ in the shown $Ta$ regime.
\label{fig:07}}
\end{figure}

A three-dimensional overview of the found parameter dependences of  the angular velocity transport $Nu_\omega (Ta,a)$ is shown in figure \ref{fig:06}. One immediately observes a pronounced maximum in $Nu_\omega (Ta,a)$ with a considerable offset from the line $a=0$. A more detailed view is obtained in cross-sections through figure \ref{fig:06} and in particular in {\it compensated plots} as shown in figure \ref{fig:07}(\emph{a}), where we divided $Nu_\omega$ by the approximate effective scaling $\sim Ta^{0.39}$. In this way we identify a universal effective scaling $Nu_\omega (Ta, a)\propto Ta^{0.39}$ by averaging over the complete $Ta$-range, ignoring $Ta$-dependence and thus calling the scaling effective. If each curve for each $a$ is fitted individually, the resulting $Ta$ scaling exponents $\gamma(a)$ scatter with $a$, but at most very slightly depend on $a$; see figure \ref{fig:07}(\emph{b, c}). For different linear fits below and above $\aOpt = 0.33$ (actually below and above $a_{bis} = 0.368$ or $\psi = 0$, as will be introduced later on), we obtain $\gamma = 0.378 + 0.028 a \pm 0.01$ for $a<\aOpt$, the exponent slightly decreasing towards less counter-rotation, and a constant exponent $\gamma = 0.394 \pm 0.006$ for increasing counter-rotation beyond the optimum. The trend in the exponents for $a < \aOpt$ is small and  compatible with a constant $\gamma = 0.39$ and a merely statistical scatter of $\pm 0.03$. It is in this approximation that $Nu_\omega (Ta, a)$ factorizes.

\begin{landscape}
\begin{table}
\setlength{\tabcolsep}{3.6pt}
\begin{tabular}{rrrrrrrrrrrrrrrrrr}
$a$   & $\psi$ & $Ta$  & $Ta$  & $\omega_i$ & $\omega_i$ & $\omega_o$ & $\omega_o$ & $Re_i$ & $Re_i$ & $Re_o$ & $Re_o$ & $G$   & $G$   & $Nu_{\omega}$ & $Nu_{\omega}$ & $\gamma(a)$ & $f(a)$ \\
 & & min & max & min & max & min & max & min & max & min & max & min & max & min & max & &  \\
 & (deg.) & $(10^{11})$ & $(10^{11})$ & (rad s$^{-1}$) & (rad s$^{-1}$) & (rad s$^{-1}$) & (rad s$^{-1}$) & $(10^5)$ & $(10^5)$ & $(10^5)$ & $(10^5)$ & $(10^9)$ & $(10^9)$ &       &  & & $(10^{-3})$ \\
 \\
2.000  & 43.1  & 4.07  & 34.0  & 10.5  & 30.7  & -61.5 & -21.1 & 1.72  & 4.99  & -13.95 & -4.82 & 0.49  & 3.29  & 91    & 212 & 0.397 & 2.71\\
1.000  & 27.2  & 2.07  & 62.3  & 11.3  & 61.9  & -61.9 & -11.3 & 1.84  & 10.12 & -14.15 & -2.58 & 0.49  & 10.30 & 129   & 492 & 0.386 & 4.84\\
0.714  & 17.7  & 1.52  & 57.1  & 11.4  & 69.3  & -49.5 & -8.1  & 1.84  & 11.31 & -11.29 & -1.84 & 0.47  & 12.08 & 143   & 603 & 0.399 & 6.21\\
0.650  & 15.0  & 1.85  & 52.3  & 12.0  & 62.6  & -40.7 & -7.8  & 2.12  & 11.25 & -10.21 & -1.92 & 0.58  & 11.91 & 162   & 621 & 0.396 & 6.58\\
0.600  & 12.8  & 1.52  & 52.2  & 11.4  & 65.8  & -39.5 & -6.9  & 1.97  & 11.59 & -9.71 & -1.65 & 0.53  & 12.57 & 162   & 656 & 0.392 & 6.98\\
0.550  & 10.3  & 1.63  & 53.7  & 11.9  & 67.4  & -37.1 & -6.5  & 2.11  & 12.13 & -9.33 & -1.63 & 0.57  & 13.42 & 168   & 691 & 0.401 & 7.23\\
0.500  & 7.7   & 1.43  & 57.5  & 11.7  & 73.3  & -36.7 & -5.9  & 2.04  & 12.97 & -9.06 & -1.43 & 0.55  & 15.32 & 172   & 762 & 0.398 & 7.66\\
0.450  & 4.9   & 1.33  & 66.5  & 12.0  & 83.0  & -37.4 & -5.4  & 2.04  & 14.43 & -9.08 & -1.28 & 0.54  & 18.07 & 177   & 836 & 0.396 & 7.98\\
0.400  & 2.0   & 1.15  & 85.4  & 12.6  & 93.6  & -37.5 & -5.0  & 1.97  & 16.93 & -9.47 & -1.10 & 0.52  & 22.96 & 184   & 937 & 0.386 & 8.36\\
0.368  & 0.0   & 2.65  & 63.2  & 18.9  & 89.2  & -32.8 & -7.0  & 3.05  & 14.90 & -7.67 & -1.57 & 1.08  & 18.20 & 250   & 864 & 0.389 & 8.60\\
0.350  & -1.2  & 1.16  & 64.6  & 12.3  & 90.1  & -31.5 & -4.3  & 2.04  & 15.28 & -7.47 & -1.00 & 0.52  & 18.67 & 181   & 876 & 0.391 & 8.61\\
0.300  & -4.5  & 1.15  & 65.0  & 12.4  & 93.0  & -27.9 & -3.7  & 2.11  & 15.91 & -6.67 & -0.89 & 0.52  & 18.36 & 185   & 859 & 0.383 & 8.60\\
0.250  & -8.0  & 1.07  & 63.4  & 13.0  & 97.5  & -24.4 & -3.2  & 2.12  & 16.35 & -5.71 & -0.74 & 0.48  & 17.35 & 177   & 822 & 0.381 & 8.37\\
0.200  & -11.6 & 2.03  & 67.7  & 19.0  & 105.8 & -21.2 & -3.8  & 3.05  & 17.59 & -4.91 & -0.85 & 0.83  & 17.57 & 219   & 805 & 0.375 & 8.07\\
0.143  & -15.9 & 2.27  & 69.4  & 22.1  & 112.5 & -16.1 & -3.2  & 3.38  & 18.70 & -3.73 & -0.68 & 0.83  & 17.21 & 208   & 779 & 0.386 & 7.69\\
0.100  & -19.3 & 4.78  & 60.0  & 31.5  & 112.6 & -11.3 & -3.2  & 5.10  & 18.08 & -2.52 & -0.71 & 1.56  & 14.72 & 269   & 717 & 0.395 & 7.37\\
0.000  & -27.2 & 1.34  & 61.7  & 18.2  & 124.0 & 0.0   & 0.0   & 2.97  & 20.15 & 0.00  & 0.00  & 0.49  & 13.73 & 158   & 660 & 0.375 & 6.81\\
-0.140 & -38.3 & 1.90  & 37.4  & 25.5  & 112.4 & 3.6   & 15.7  & 4.11  & 18.25 & 0.80  & 3.57  & 0.55  & 7.07  & 151   & 436 & 0.364 & 5.76\\
-0.200 & -42.8 & 2.08  & 29.6  & 28.6  & 107.7 & 5.7   & 21.5  & 4.62  & 17.44 & 1.29  & 4.87  & 0.49  & 5.10  & 128   & 354 & 0.392 & 5.00\\
-0.400 & -56.4 & 4.87  & 22.2  & 58.3  & 124.3 & 23.3  & 49.8  & 9.44  & 20.16 & 5.28  & 11.27 & 0.47  & 1.68  & 80    & 135 & 0.358 & 2.21\\

\end{tabular}
\caption{The measured global torque data for the individual cases of fixed $a\equiv-\omega_o/\omega_i$ at increasing $Ta$, as presented in figure \ref{fig:05}, equivalent to straight lines in the ($Re_o$, $Re_i$) parameter space, as presented in figure \ref{fig:04}(\emph{a}). We list the minimum and maximum values of the driving parameters ($Ta$, $\omega_i$, $\omega_o$, $Re_i$, and $Re_o$) and we list the minimum and maximum values of the response parameters (dimensionless torque $G(Ta,a)$ and dimensionless angular velocity transport flux $Nu_{\omega}(Ta,a) = f(a) F(Ta)$). The variable $a$ can be transformed to the angle $\psi$, i.e.\ the angle in ($Re_o$, $Re_i$) space between the straight line characterized by $a$ and that characterized by $a_{bis}$, describing the angle bisector of the instability range; see figure \ref{fig:04}(\emph{a}) and (\ref{eq:psi}). The second to last column lists the effective scaling exponent $\gamma(a)$ obtained from fitting $Nu_{\omega}(Ta, a)$ to a least-squares linear fit in log-log space for each individual case of $a$, as presented in figure \ref{fig:07}(\emph{b, c}). The prefactor $f(a)$, as given in the last column, is determined by fixing $\gamma(a)$ at its average encountered value of $\gamma\approx0.39$ and by averaging the compensated $Nu_{\omega}Ta^{-0.39}$ over $Ta$, as presented in figure \ref{fig:10}.
\label{tab:G-data}}
\end{table}
\end{landscape}

\clearpage

\subsection{Ultimate regime}\label{ult}

\Citet{gil11} interpreted the effective scaling $Nu_\omega(Ta, a) \sim Ta^{0.38}$, similar to our currently obtained $\gamma=0.39\pm0.03$, as an indication of the so-called `ultimate regime' -- distinguished by both turbulent bulk and turbulent boundary layers. Such scaling was predicted by \cite{gro11} for very strongly driven RB flow. As detailed in \cite{gro11}, it emerges from a $Nu_\omega (Ta ) \sim Ta^{1/2}$ scaling with logarithmic corrections originating from the turbulent boundary layers. Remarkably, the corresponding wind Reynolds number scaling in RB flow does {\em not} have logarithmic corrections, i.e.\ $Re_w \sim Ta^{1/2}$. These RB scaling laws for the thermal Nusselt number and the corresponding wind Reynolds number have been confirmed experimentally by \cite{ahl11a} for $Nu$ and by \cite{he12} for $Re_w$. According to the EGL theory this should have its correspondence in TC flow. That leads to the interpretation of $\gamma = 0.39$ as an indication for the ultimate state in the presently considered TC flow. Furthermore, \cite{hui12} indeed also found from particle image velocimetry (PIV) measurements in the present strongly driven TC system the predicted \citep{gro11} scaling of the wind, $Re_w \propto Ta^{1/2}$.

We note that in our available $Ta$ regime the effective scaling law $Nu_\omega  \sim Ta^{0.39}$ is practically indistinguishable from the prediction of \cite{gro11}, namely,
\be
Nu_{\omega}\sim Ta^{1/2} \mathcal{L}(Re_w(Ta)),
\label{nu-log-cor}
\ee
with the logarithmic corrections $\mathcal{L}(Re_w(Ta))$ detailed in equations (7) and (9) of \cite{gro11}. The result from (\ref{nu-log-cor}) is shown as a solid line in figure \ref{fig:07}(\emph{a}), showing the compensated plot $Nu_\omega/Ta^{0.39}$. Indeed, only detailed inspection reveals that the theoretical line is not exactly horizontal.

\begin{figure}
\centerline{\includegraphics[width=1\textwidth]{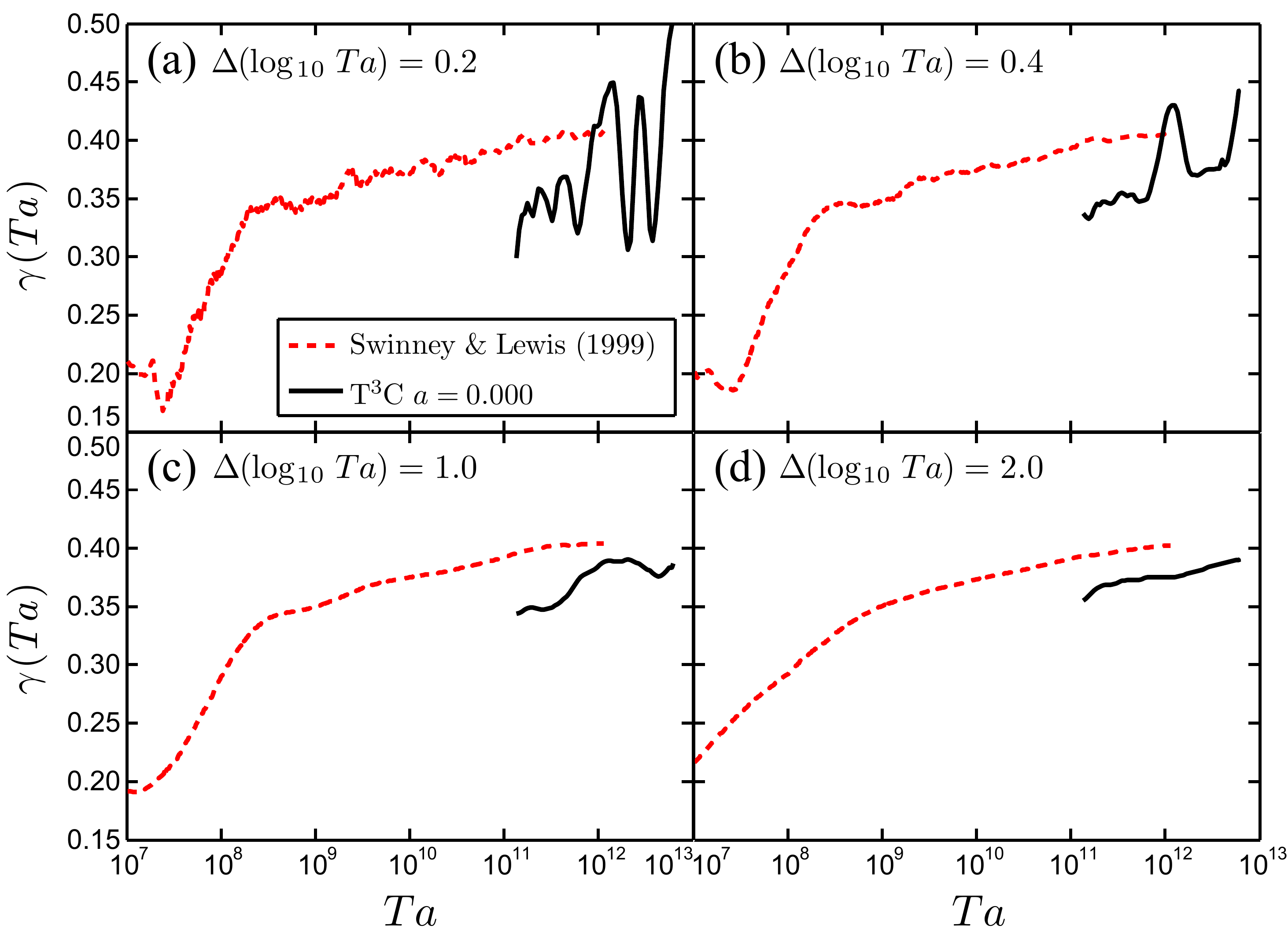}}
\caption{Local $Ta$-dependent scaling exponent $\gamma(Ta) = d(\mathrm{log_{10}} Nu_{\omega})/d(\mathrm{log_{10}}Ta)$ for the case of inner cylinder rotation only ($a = 0$). The black solid line is our experimental data and the dark grey (red) dashed line is data from \cite{lew99}, see their figure 3 but now transformed into ($Ta, Nu_{\omega}$) space. The local exponent is calculated by using a sliding least-squares linear fit over different intervals: (a) $\Delta(\mathrm{log_{10}}Ta) = 0.2$, (b) 0.4, (c) 1.0 and (d) 2.0. This method is similar to the that used in \cite{lew99}. Our data reveal a detailed local sensitivity of the scaling exponent on small $Ta$ intervals; see (a). When fitting over wider $Ta$ intervals, an overall increasing $\gamma(Ta)$ with $Ta$ becomes apparent; see (d).
\label{fig:08}}
\end{figure}

Thus, strictly speaking, there is a $Ta$ dependence of the scaling exponent $\gamma(Ta)$, as was clearly evidenced by \cite{lat92a} and \cite{lew99} in a much larger $Ta$ range. In figure \ref{fig:08} we present our local $\gamma(Ta)$ for the case of $a = 0.000$ and we compare it to the data from \cite{lew99}. Similar to \cite{lew99} we calculate $\gamma(Ta)=d(\mathrm{log_{10}}Nu_{\omega})/d(\mathrm{log_{10}}Ta$) by using a sliding least-squares linear fit over a certain $\Delta(\mathrm{log_{10}}Ta)$ range, as indicated by the top left corner of panels (\emph{a--d}). The narrow averaging range used in figure \ref{fig:08}(\emph{a}) results into a strongly fluctuating $\gamma(Ta)$. The origin of these fluctuations may be  different turbulent flow states (e.g.\ different number of Taylor vortices); future studies should shed more light onto this. When averaging over a wider $Ta$ range, our data recover  a monotonically  increasing $\gamma(Ta)$ trend, as can be seen in figure \ref{fig:08}(\emph{d}), which is in line with \cite{lat92a} and \cite{lew99}. Clearly, with their large-$Ta$ measurements, these authors also already were in the ultimate TC regime.

\begin{figure}
\centerline{\includegraphics[width=1\textwidth]{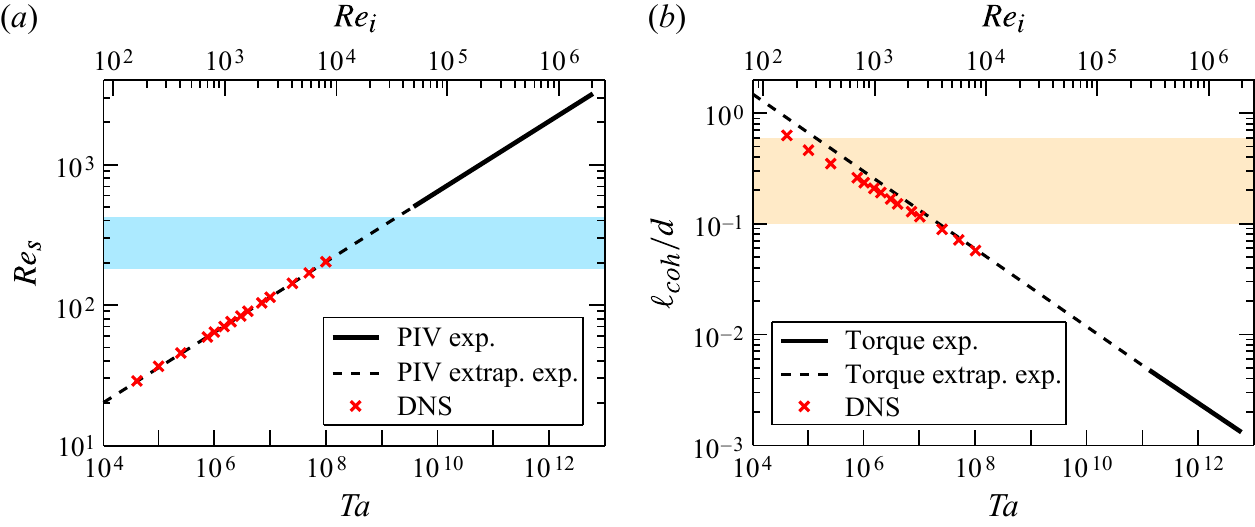}}
\caption{(a) The shear Reynolds number $Re_s$, and (b) the coherence length $\ell_{coh}$, estimated as 10 times the Kolmogorov length scale $\eta_K$, over the TC-gap width $d$, both  versus the driving strength ($Ta$, lower abscissa; $Re_i$, upper abscissa) for the case of pure inner cylinder rotation ($a=0$). (a) The solid black line results from the experimental data obtained by PIV \citep{hui12}. The extrapolation of these data (dashed black line) towards smaller $Ta$ nicely agrees with the DNS data (grey/red crosses) of \cite{ost12}. The (pale blue) shaded area indicates the transitional regime from moderate turbulence at lower $Ta$ (turbulent bulk with laminar BLs) into ultimate turbulence at higher $Ta$ (turbulent bulk with turbulent BLs). (b) The solid black line is experimental data obtained by global torque measurements (this work). The extrapolation of these data (dashed black line) towards smaller $Ta$ agrees with the DNS data (grey/red crosses) of \cite{ost12}. The (pale yellow) shaded area indicates the transitional regime where spatial coherence becomes small enough to allow for a turbulent bulk beyond this regime.
\label{fig:09}}
\end{figure}

This gives rise to the following question: Where does the ultimate turbulence regime set in for turbulent TC flow? To find out, we calculate the shear Reynolds number $Re_s = U_s \delta /\nu$, where $\delta$ is the thickness of the kinetic boundary layer, still being of Prandtl type, and $U_s$ is the shear velocity across $\delta$. The latter we estimate as $U_s = U_i - U_w$. Correspondingly, we estimate the kinetic Prandtl-Blasius type BL thickness as $\delta = a_{PB}d/\sqrt{Re_i - Re_w}$ (see e.g.\ \cite{ll87}), with $a_{PB}$ set to $2.3$.
This results in a shear Reynolds number of  $Re_s=a_{PB}\sqrt{Re_i - Re_w}$.
For the wind Reynolds number we take our experimental  result based on PIV measurements
(\cite{hui12}), namely $Re_w = 0.0424Ta^{0.495}$ (in the $Ta$ regime from $3.8\times10^{9}$ to $6.2\times10^{12}$, for $a=0$). This implies that the relative contribution of the wind $Re_w/Re_i = U_w/U_i$ is only around $4.6\%$ in this regime. Nonetheless, we take it into consideration in figure \ref{fig:09}(\emph{a}), in which we plot $Re_s$ versus $Ta$, retrieving the effective scaling $Re_s = 2.02 Ta^{0.25}$. That figure also shows the result of \cite{ost12} from direct numerical simulation (DNS), who found $Re_w = 0.0158Ta^{0.53}$ in the $Ta$ regime from $4\times10^4$ to $1\times10^7$. Again, also here  the relative contribution of the wind $Re_w/Re_i$ is very small, namely around $3\%$. These numerical results give an  effective scaling of $Re_s = 2.05 Ta^{0.25}$, very similar to our experimental findings, even prefactor-wise.

The Prandtl-Blasius type BL becomes turbulent for a shear Reynolds number larger than a critical shear Reynolds number or transition shear Reynolds number $Re_{s,T}$, which is known to be in the range between 180 and 420 (see e.g.\ \cite{ll87}). This range is shown as shaded in figure \ref{fig:09}: all of our experimental data points of this present paper (solid line) are beyond that onset.  So, indeed, we are in the ultimate regime. In contrast, the numerical data points by \cite{ost12} are in the Prandtl-Blasius regime with  laminar-type boundary layers.

The transition between these two regimes occurs in between. The range 180--420 for the
transitional shear Reynolds number $Re_{s,T}$ here (i.e., for the present $\eta$ and $a=0$) corresponds to a range between $3 \times 10^7$ and $10^9$ for the transitional Taylor number  $Ta_T$ and to a range between $5 \times 10^3$ and $2 \times 10^4$ for the transitional (inner) Reynolds number $Re_{i,T}$. This corresponds to the transitional  Reynolds number found by \cite{lew99}, see their figure 3, in which the transition to the ultimate regime is identified at a Reynolds number $Re_{i,T} = 1.3 \times 10^4$. Below that value \cite{lew99} find a very steep increase of the local slope $d\log Nu_\omega / d\log Re_i$ with $Re_i$; beyond the transition the increase is much less. (Here we have translated \cite{lew99}'s finding into the notation of this present paper.) We stress again that the values given in this and the next subsection hold for $a=0$. How the values of the transitional Reynolds or Taylor number depend on $a$ remains an important question for future research.

Both in our experiment and in the experiments by \cite{lew99} the logarithmic corrections
in (\ref{nu-log-cor}) are visible and have the consequence that the ``real'' ultimate scaling $Nu_\omega \sim Ta^{1/2} $ is never achieved. As explained in \cite{gro11} -- and, differently and with a different result, much earlier in \cite{kra62}) -- these logarithmic corrections are a consequence of the logarithmic velocity profile in the turbulent boundary layers. Only by destroying these logarithmic profiles by extreme wall roughness as done in TC experiments by \cite{ber03} or in RB experiments by \cite{roc01} or by replacing the walls by periodic boundary conditions (and a volume forcing) as done in numerical simulations by \cite{loh03,cal05,sch12} can one recover the 1/2 scaling exponent, which is obtained in  the strict upper-bound of \cite{doe94}.

\subsection{Comparison Taylor-Couette turbulence with Rayleigh-B\'enard turbulence}

To get an idea of the extension of the non-ultimate turbulence regime in TC flow, we also estimate the coherence length $\ell_{coh}$, below which the spatial coherence of structures in the flow becomes small enough to allow for developed turbulence in the flow. Typically, one estimates the coherence length as a multiple of the (mean) Kolmogorov length scale $\eta_K = \nu^{3/4}/\epsilon^{1/4}$, namely  $\ell_{coh} \approx 10 \eta_K$. The factor of 10 between these two length scales is motivated by the transition between viscous subrange and inertial subrange, which is known to happen at a scale around $10\eta_K$ (see e.g.\ \cite{eff87}). Here $\epsilon$ is the mean energy dissipation rate. That can be obtained from the angular velocity flux $J^\omega$ (see (4.7) of EGL), namely
\be
\epsilon =
{2 (\omega_i - \omega_o) J^\omega \over r_o^2 - r_i^2 } =
{2 (\omega_i - \omega_o) J^\omega_{lam} Nu_\omega \over r_o^2 - r_i^2 },
\label{egl4.7}
\ee
reflecting the statistical balance between external driving and internal dissipation. As pointed out in EGL, a more elegant way to write this balance is
\be
\epsilon - \epsilon_{lam} = {\nu^3 \over d^4} (Nu_\omega - 1) Ta \sigma^{-2}.
\label{egl4.18}
\ee
In any case, we can use our data for $Nu_\omega$ to calculate the mean Kolmogorov length scale
$\eta_K$ and thus the coherence length $\ell_{coh} = 10 \eta_K$. In figure \ref{fig:09}(\emph{b}) we show $\ell_{coh}$ as a function of $Ta$ or $Re_i$ (for $a=0$ and $\eta = 0.716$). When the  coherence length becomes  smaller than, say, 0.1--0.5 times the outer length scale $d$, one can reasonably start to speak of a developed turbulence regime in the bulk in between the inner scale $\ell_{coh} = 10\eta_K$ and the outer scale $d$. According to figure \ref{fig:09}(\emph{b}), this onset of a developed turbulence regime in the bulk, but still with Prandtl-Blasius type boundary layers (see \cite{sun08, zho10b}), occurs at Taylor numbers  $Ta_{on}$ between $2\times 10^5$ and $2\times 10^7$ or onset (inner) Reynolds numbers $Re_{i,on}$ between 300 and 3000, far below the regime of our present experiments, but  in the regime of the numerical simulations of \cite{ost12}. The corresponding numbers for smaller coherence in RB flow are given in figure 1 of  \cite{sug07}. For $Pr\simeq 1$, a developed turbulence regime in the bulk becomes possible beyond $Ra \simeq 10^7$.

\begin{table}
\begin{center}
\begin{tabular}{ l c c }
&  TC
&  RB
\\
\\
Loss of spatial coherence
&  $Ta_{on} \simeq 10^6 $
&  $Ra_{on} \simeq 10^7  $
\\
(defined via $\ell_{coh}$)
&  $Re_{i,on} \simeq 10^3$
&
\\
\\
BL shear instability
&  $Ta_T \simeq 5\times 10^8$
&  $Ra_T \simeq 10^{14}$
\\
(defined via $Re_s$)
&   $Re_{i,T} \simeq  10^{4}$
&
\\
 \end{tabular}
 \end{center}
\caption{
Estimates for the onset of the regime where a turbulent bulk becomes possible next to laminar-type BLs (upper part) and for the transition towards the ultimate regime in which the BLs are turbulent (lower part), for both TC and RB turbulence. For TC the estimates are derived here. For RB they are taken from the literature: from \cite{sug07} for $Ra_{on}$ (for $Pr\simeq 1$); and from \cite{gro01} (theoretical prediction) and from \cite{he12} (experimental confirmation) for $Ra_T$. In between these values laminar-type BLs and a turbulent bulk can coexist and the transport properties can be described by the unifying RB theory of \cite{gro00,gro01,gro02,gro04}, which has been extended to TC by EGL. Beyond $Ta_T$ (respective $Ra_T$) the turbulent nature of BLs
lead to different scaling properties, as elaborated in \cite{gro11}.
}
\label{tab-onsets}
\end{table}

Figure \ref{fig:09}(\emph{a, b}) reveals that there should be a TC flow range with Prandtl-Blasius type (laminar) BLs and a turbulent bulk roughly in between $Ta \simeq 10^6$ and $Ta \simeq 5\times 10^8$ or $Re_i$ in between $Re_i \simeq 10^3 $ and $10^4$. This regime was explored in the earlier experiments by \cite{lat92,lat92a,lew99} and others -- it cannot be accessed with water as operating liquid in our T$^3$C set-up as the angular velocities would have to be too low for reasonable precision. We could, however, explore that regime with more viscous liquids also with our T$^3$C set-up.

Our present estimates for TC flow and the earlier findings and estimates for RB flow are summarized in table \ref{tab-onsets}. The table gives rise to the interesting question: Why is the ``classical regime'' (as it is called by Ahlers) in between $Ra_{on}$ and $Ra_T$, in which a laminar-type BL and a turbulent bulk coexist and in which the unifying theory of \cite{gro00,gro01,gro02,gro04} is applicable, so extended in RB turbulence, but so small in TC turbulence? Or, in other words: Why does the ultimate regime with its turbulent BLs set in for much smaller $Ta$ in TC flow as compared to the extremely high $Ra$ values for that onset in RB flow?

We think that the answer lies in the much higher efficiency of the shear driving in TC flow
as compared to the thermal driving in RB flow. In RB flow the shear instability of the kinetic BL is induced by the thermal driving only indirectly; namely, the driving first induces a large scale wind, which then in turn builds up the shear near the boundaries. In TC flow the flow is directly driven by the rotating inner cylinder, giving rise to a very large direct shear. As pointed out above, the large scale wind with its strength $Re_w$ only means a small correction of $4-5\%$ to $Re_i$ in the calculation of the shear Reynolds number. As roughly $Re_s \sim Re^{1/2}  \sim Ta^{1/4}$, a factor of 20 in the shear Reynolds number leads to the huge difference $20^4=1.6\times10^5$ in the typical onset Taylor number.

\subsection{Optimal angular velocity transport}

\begin{figure}
\centerline{\includegraphics[width=1\textwidth]{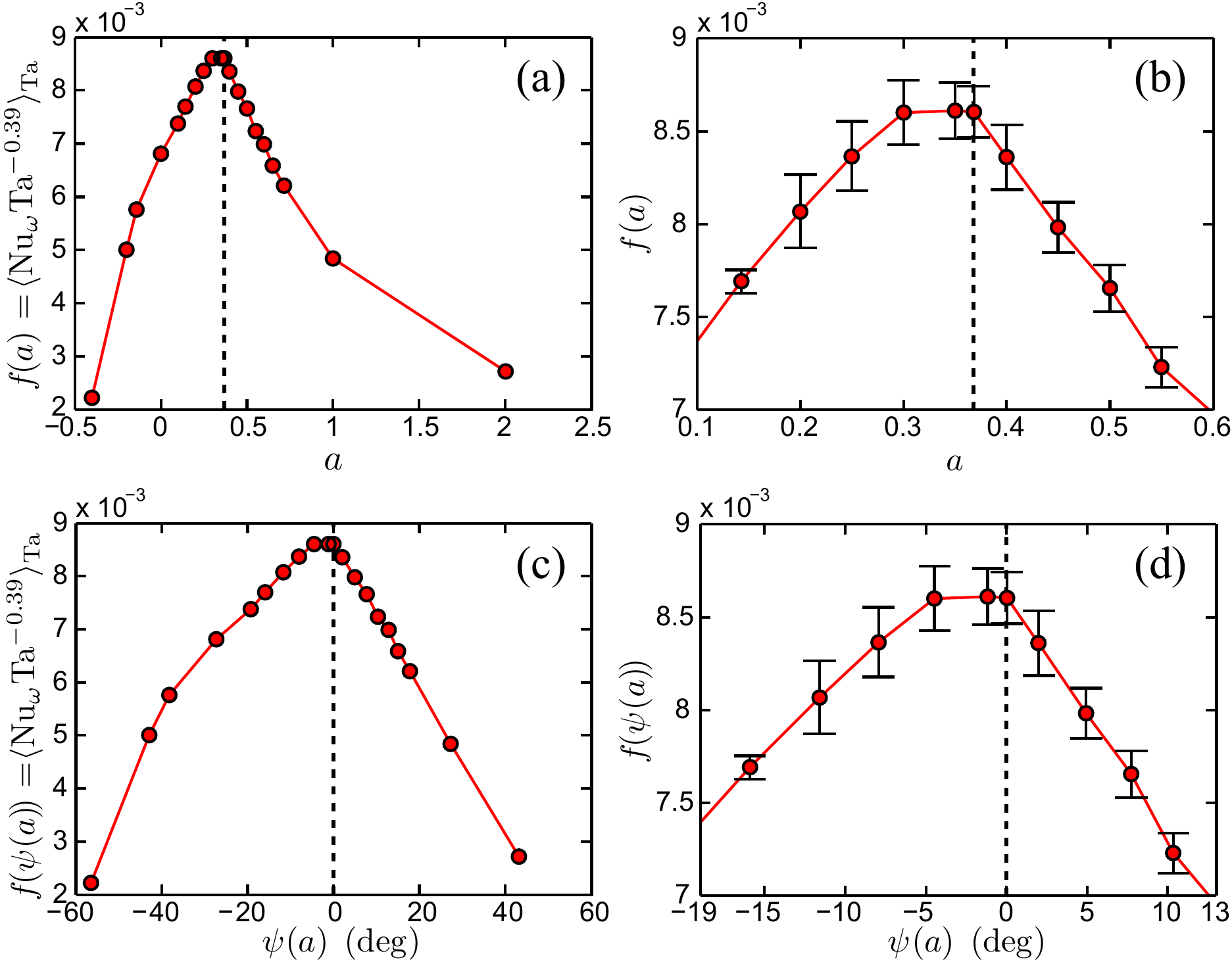}}
\caption{Amplitude of the effective scaling law, $f=\langle Nu_{\omega} Ta^{-0.39}\rangle_{Ta}$ (shown in figure \ref{fig:07}), (\emph{a, b}) as function of $a$ and (\emph{c, d}) as a function of $\psi(a)$. The dashed line in all panels corresponds to the suggested case of optimal turbulence as given by the angle bisector (\ref{eq:a_bis}), i.e.\ $a_{bis}=0.368$. (The connecting lines between the data points are guides for the eyes.) The standard deviation of $Nu_{\omega}Ta^{-0.39}$ is similar to the size of the symbols in (\emph{a}) and (\emph{c}), and is indicated by the error bars of the zoomed-in panels (\emph{b}) and (\emph{d}). The angular velocity transport flux amplitude is systematically larger towards the co-rotating instability borders $a = -\eta^2$ or $\psi \approx -62.8^\circ$ with respect to the counter-rotating instability borders $a\rightarrow\infty$ or $\psi \approx 62.8^\circ$.
\label{fig:10}}
\end{figure}

Coming back to our experimental data on TC: the (nearly) horizontal lines in figure \ref{fig:07}(\emph{a}) imply that $Nu_{\omega}(Ta, a)$ within the present experimental precision nearly factorizes in $Nu_{\omega}(Ta, a) = f(a) Ta^{0.39 \pm 0.03}$. We now focus on the $a$ dependence of the angular velocity flux amplitude $f(a) = Nu_{\omega}(Ta, a) / Ta^{0.39}$, shown in figure \ref{fig:10}(\emph{a, b}), and its interpretation.  One observes a very pronounced maximum at $\aOpt = 0.33 \pm 0.04$, reflecting the optimal angular velocity transport from the inner to the outer cylinder at that angular velocity ratio. This value is obtained by averaging over the three data points making up the small plateau visible in figure \ref{fig:10}(\emph{b}). Naively, one might have expected that $f(a)$ has its maximum at $a=0$, i.e.\ $\omega_o = 0$ (no outer cylinder rotation) since outer cylinder rotation stabilizes an increasing part of the flow volume for increasing counter-rotation rate. On the other hand, outer cylinder rotation also enhances the total shear of the flow, leading to enhanced turbulence, and thus more angular velocity transport is expected. The $a$ dependence of $f(a)$ thus reflects the mutual importance of both these effects.

\begin{figure}
\centerline{\includegraphics[width=1\textwidth]{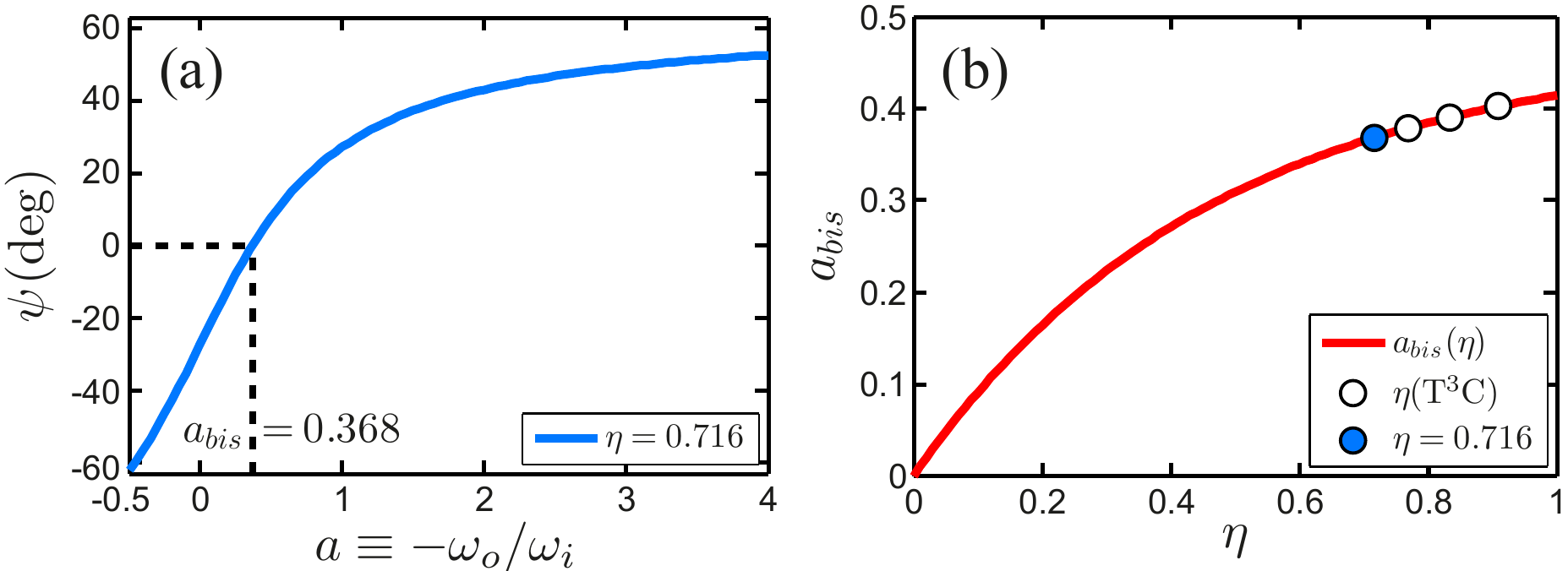}}
\caption{(\emph{a}) The transformation from $a$ to $\psi$ as given by (\ref{eq:psi}), shown here for the radius ratio $\eta=0.716$ used in the present work. Note that the domain of $a = [-\eta^2, \infty]$ from co- to counter-rotation, spanning the complete unstable flow regime, is transformed to the range $\psi\approx[-62.80^\circ, + 62.80^\circ]$ for this specific $\eta$. (\emph{b}) The dependence of $a_{bis}$ on $\eta$ as given by (\ref{eq:a_bis}), shown as the grey (red) line. The four circles show the radius ratios accessible in the T$^3$C, i.e.\ $\eta = 0.716, 0.769, 0.833, 0.909$. The grey (blue) filled circle is the radius ratio $\eta = 0.716$ of the present work, suggesting optimal turbulence at $a_{bis} = 0.368$.
\label{fig:11}}
\end{figure}

Generally, one expects an increase of the turbulent transport if one goes deeper into the control parameter range ($Ta, a$) in which the flow is unstable. We speculate that the optimum positions for the angular velocity transport should consist of all points in parameter space that are equally distant from both the right branch (first quadrant, co-rotation) of the instability border and its left branch (second quadrant, counter-rotation). In the inviscid approximation ($\nu = 0$), these two branches are given by the Rayleigh criterion, $\partial(r^2 \omega)/\partial(r)>0$, resulting in the lines given by the relations $\omega_i/\omega_o|_\mathrm{Rayleigh}=\eta^{-2}$ and $\omega_i=0$, which translate into $a = - \eta^2$ and $a = \infty$, respectively. The line of equal distance from both is the angle bisector of the instability range. Its relation can easily be calculated to be
\begin{equation}
a_{bis}(\eta) = \frac{\eta}{\mathrm{tan}\left[\frac{\pi}{2}-\frac{1}{2}\mathrm{arctan}\left(\eta^{-1}\right)\right]}.
\label{eq:a_bis}
\end{equation}
For $\eta = 0.716$ this gives $a_{bis} = 0.368$. Noteworthily, the measured value $\aOpt = 0.33 \pm 0.04$ agrees indistinguishably within experimental precision with the bisector line, supporting our interpretation. It also explains why only the lines $a =$ const scan the parameter space properly. This reflects the straight character of the linear instability lines -- as long as one is not too close to them to see the details of the viscous corrections, i.e.\ if $Ta$ is large and $a$ is well off the instability borders at $a = - \eta^2$ for co-rotation and $a \rightarrow \infty$ for counter-rotation.

Instead of characterizing the lines by the slope parameter $a$, one can introduce the angle $\psi$ between the line of chosen $a$ and the angle bisector of the instability range denoted by $a_{bis}$; thus $a = a_{bis}$ corresponds to $\psi = 0$:
\begin{equation}
\psi(a) = \frac{\pi-\mathrm{arctan}\left(\eta^{-1}\right)}{2}-\mathrm{arctan}\left(\frac{\eta}{a}\right).
\label{eq:psi}
\end{equation}
The transformation (\ref{eq:psi}) is shown in figure \ref{fig:11}(\emph{a}) and the resulting $f(\psi(a))$ in figure \ref{fig:10}(\emph{c}). The function $f(a)$ as a function of $a$ is strongly asymmetric both around its peak at $\aOpt$ and at its tails, presumably because of the different viscous corrections at $a = -\eta^2$ (decreasing $\propto Re_o^{-2}$ towards the inviscid Rayleigh line) and at $a = \infty$ (non-vanishing, even increasing correction $\propto Re_o^{3/5}$, and non-normal nonlinear (shear) instability (see e.g.\ \cite{gro00rmp})).

We do not yet know whether the optimum of $f(a)$ coincides with the bisector of the Rayleigh-unstable domain for {\it all} $\eta$. Both could  coincide incidentally for $\eta = 0.716$, analysed here. But if this were the case for all $\eta$, we can predict the $\eta$ dependence of $a_{opt}(\eta)$. This, then, is given by (\ref{eq:a_bis}). This function is plotted in figure \ref{fig:11}. In future experiments we shall test this dependence with our T$^3$C facility. The three extra points we will be able to achieve are marked as white, empty circles. The precision of our facility is good enough to test (\ref{eq:a_bis}), but clearly further experiments at much smaller $\eta$ are also needed.

We note that, for smaller $Ta$, one can no longer approximate the instability border by the inviscid Rayleigh lines. The effect of viscosity on the shape of the border lines has to be taken into account. The angle bisector of the instability range in ($Re_o$, $Re_i$) parameter space (figure \ref{fig:04}(\emph{a})) will then deviate from a straight line; we therefore also expect this for $\aOpt$. The viscous corrections of the Rayleigh instability criterion were first numerically calculated by \citet{dom84} for the case of $a=0$, and then analytically estimated by \cite{ess96} and later fitted by \cite{dut07}. Figure \ref{fig:04}(\emph{b, c}) shows enlargements of the ($Re_o$, $Re_i$) parameter space, together with the Rayleigh criterion (dotted green lines) and the \cite{ess96} analytical curve (dashed-dotted black) for $\eta=0.716$. Note that the minimum of that curve is {\it not} at $Re_o=0$, but shifted to a slightly negative value $Re_o \approx -5$, where the instability sets in at $Re_i \approx 82$. If we again assume that the optimum position for turbulent transport is distinguished by equal distance to the two branches of the Esser-Grossmann curve, we obtain the red curve in figure \ref{fig:04}(\emph{b}). On the $Ta$ scale of figure \ref{fig:04}(\emph{a, c}) it is indistinguishable from a straight line through the origin and can hence be described by (\ref{eq:a_bis}). As another consequence of the viscous corrections, the factorization of the angular velocity transport flux $Nu_{\omega} = f(a) F(Ta)$ will no longer be a valid approximation in the parameter regime shown in figure \ref{fig:04}(\emph{b}). For this to hold, $Ta$ must be large and $a$ well off the instability lines. This could be tested further by choosing the parameter $a$ sufficiently large, the line approaching or even cutting the stability border for strong counter-rotation. Then the factorization property will clearly be lost.

Future low-$Ta$ experiments and/or numerical simulations for various $a$ will show how well these ideas on understanding the existence and value of $a_{opt}$, being near or equal to $a_{bis}$, are correct or deserve modification. Of course, there will be some deviations due to the coherent structures in the flow at lower $Ta$, due to the influence of the number of rolls, etc. Similarly to how in RB the $Nu$ scaling shows discontinuities, the TC scaling exponent of $Nu_{\omega}$ shows all these structures for insufficiently large $Ta$ -- see figure 3 of \citet{lew99}, for example, in which one sees how strongly the exponent $\alpha$ depends on $Re$ up to $10^4$ ($Ta$ about up to $10^8$).

\section{Local LDA angular velocity radial profiles}\label{sec4}

\begin{figure}
\centerline{\includegraphics[width=13cm]{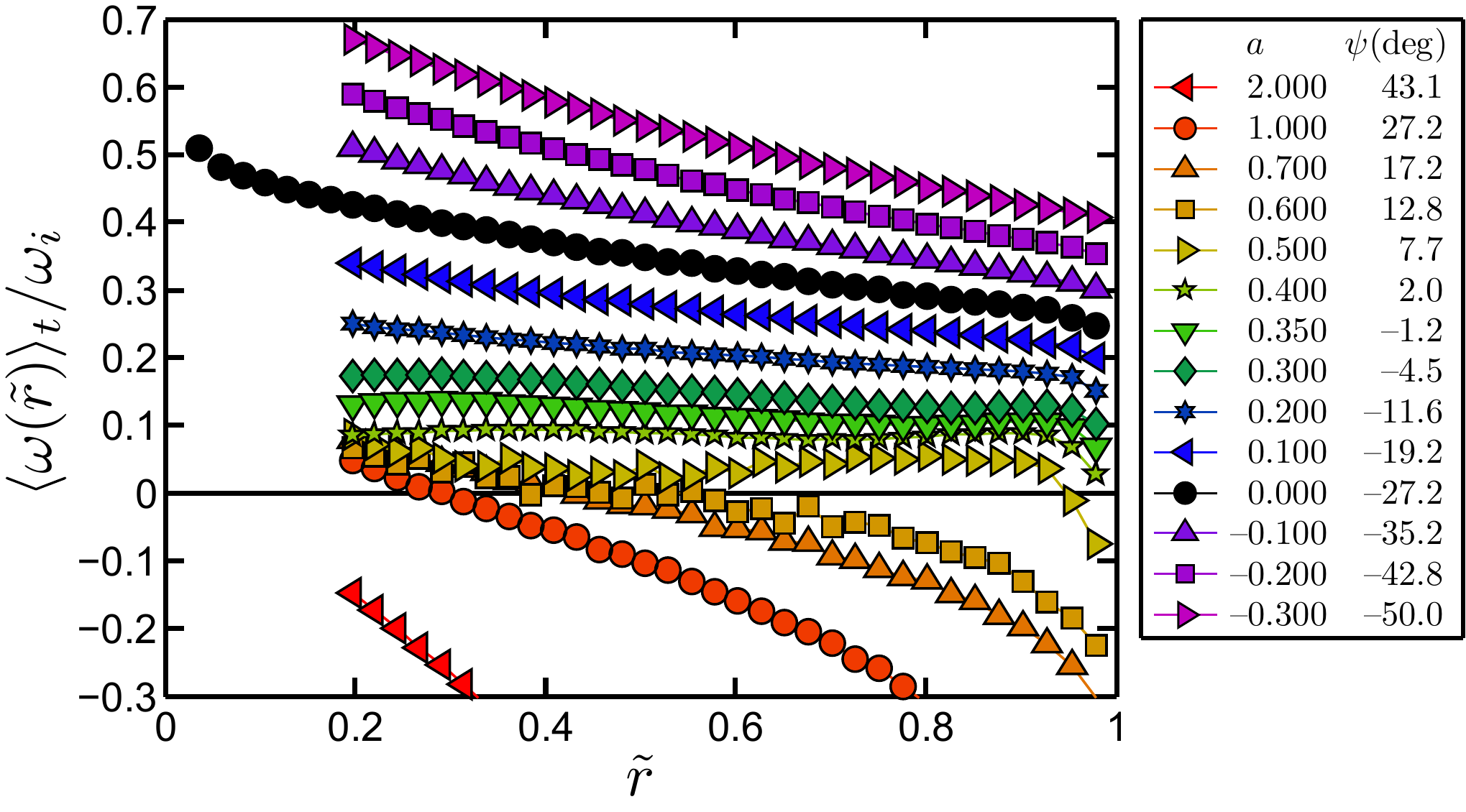}}
\caption{Radial profiles of the time-averaged angular velocity $\langle\omega(\tilde{r})\rangle_t$ at fixed height $z/L = 0.44$, normalized by the inner cylinder angular velocity $\omega_i$, for various cases of fixed $a$, as indicated by the blue filled circles in figure \ref{fig:05}. All profiles are acquired at fixed angular rotation rates of the cylinders in such a way that $Re_i - Re_o = 1.0\times 10^6$ is maintained. Instead of measuring at mid-height $z/L = 0.50$, we measure at $z/L = 0.44$, because this axial position encounters less visual obstructions located on the clear acrylic outer cylinder. To improve the visual appearance the plotted range does not fully cover the profiles corresponding to $a=1.00$ and 2.00. The profiles of $a$ = 0.50, 0.60 and 0.70 appear less smooth due to a fluctuating neutral line combined with slightly insufficient measuring time, i.e.\ convergence problems.
\label{fig:12}}
\end{figure}

\begin{figure}
\centerline{\includegraphics[width=1\textwidth]{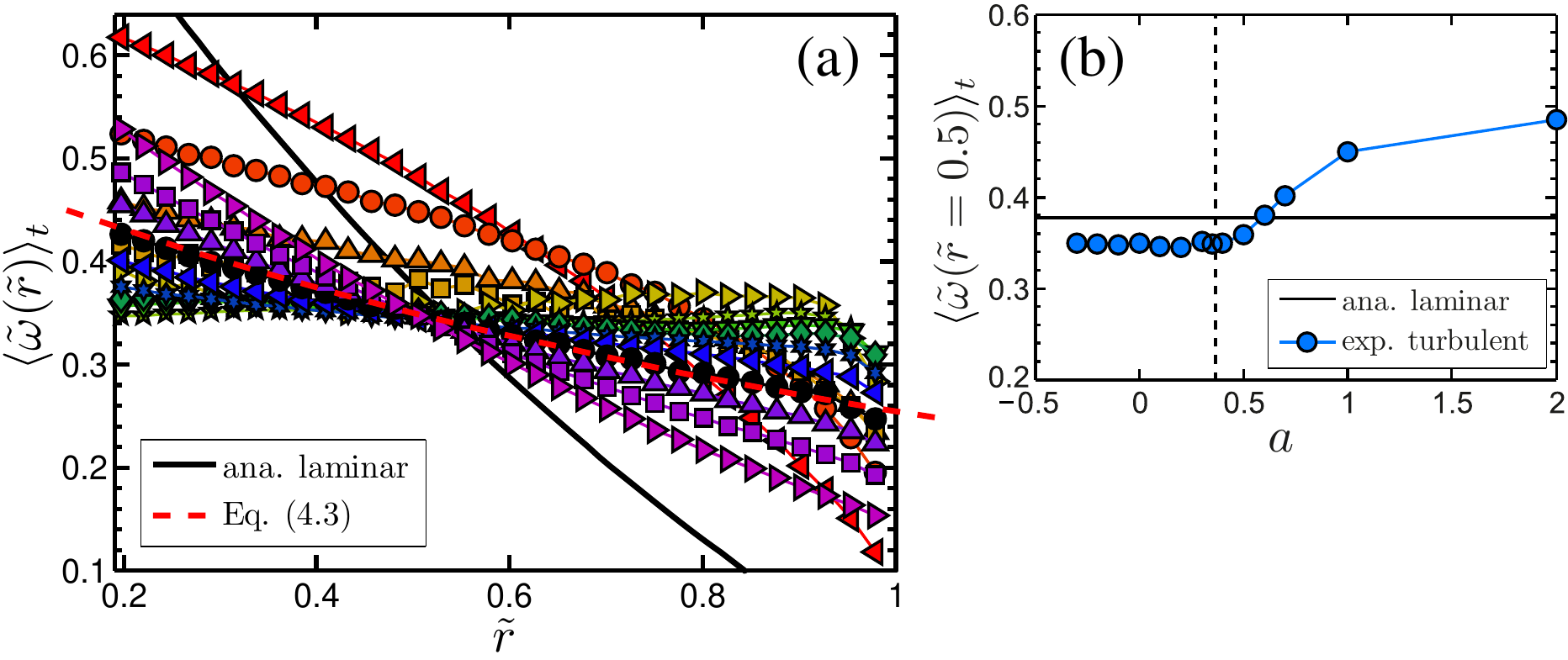}}
\caption{(\emph{a}) Rescaled angular velocity profiles $\langle \tilde{\omega}(\tilde{r}) \rangle_t = ( \langle \omega(\tilde{r}) \rangle_t - \omega_o )/(\omega_i - \omega_o)$ for various $a$. For $a \lesssim a_{bis} = 0.368$, all of these curves cross the point $(\tilde r = 1/2, \langle \tilde \omega  \rangle_t = 0.35)$. For $a > a_{bis} = 0.368$, this is no longer the case. (\emph{b}) The transition of the quantity $\left< \tilde \omega (\tilde r = 1/2 ) \right>_t$ when $a$ increases from $a\lesssim a_{bis}$ to $a_{bis} \lesssim a$. The dashed vertical line indicates $a_{bis} = 0.368 \approx  a_{opt}$.  The solid black lines in (\emph{a}) and (\emph{b}) show the $\tilde \omega (\tilde r)$ values for the laminar solution (\ref{eq:laminar_profile}). The dashed red line in (\emph{a}) shows the Busse upper-bound profile (\ref{eq:Busse2}), which is independent of the ratio $a$ and is nearly indistinguishable from the measurement for $a=0$.
\label{fig:13}}
\end{figure}

We now wonder whether the distinguishing property of the flow at $\aOpt$ (maximal angular velocity transport) is also reflected in other flow characteristics. We therefore performed LDA measurements of the  angular velocity profiles in the bulk, close to mid-height, $z/L=0.44$, at various $-0.3 \le a \le 2.0$: see table \ref{tab:LDA_experiments} for a list of all measurements, figure \ref{fig:12} for the mean profiles $\left< \omega (r,t )\right>_t$ at fixed height $z$, and figure \ref{fig:13} for the rescaled profiles $(\left< \omega (r,t )\right>_t - \omega_o )/ (\omega_i - \omega_o)$, also at fixed height $z$. With our present LDA technique, we can only resolve the velocity in the radial range $0.04\le \tilde r \le 0.98$; there is no proper resolution in the inner and outer boundary layers. Because the flow close to the inner boundary region requires substantially more time to be probed with LDA, due to disturbing reflections of the measurement volume on the reflecting inner cylinder wall necessitating the use of more stringent Doppler burst criteria, we limit ourselves to the range $0.2\le \tilde r \le 0.98$.

\begin{figure}
\centerline{\includegraphics[width=1\textwidth]{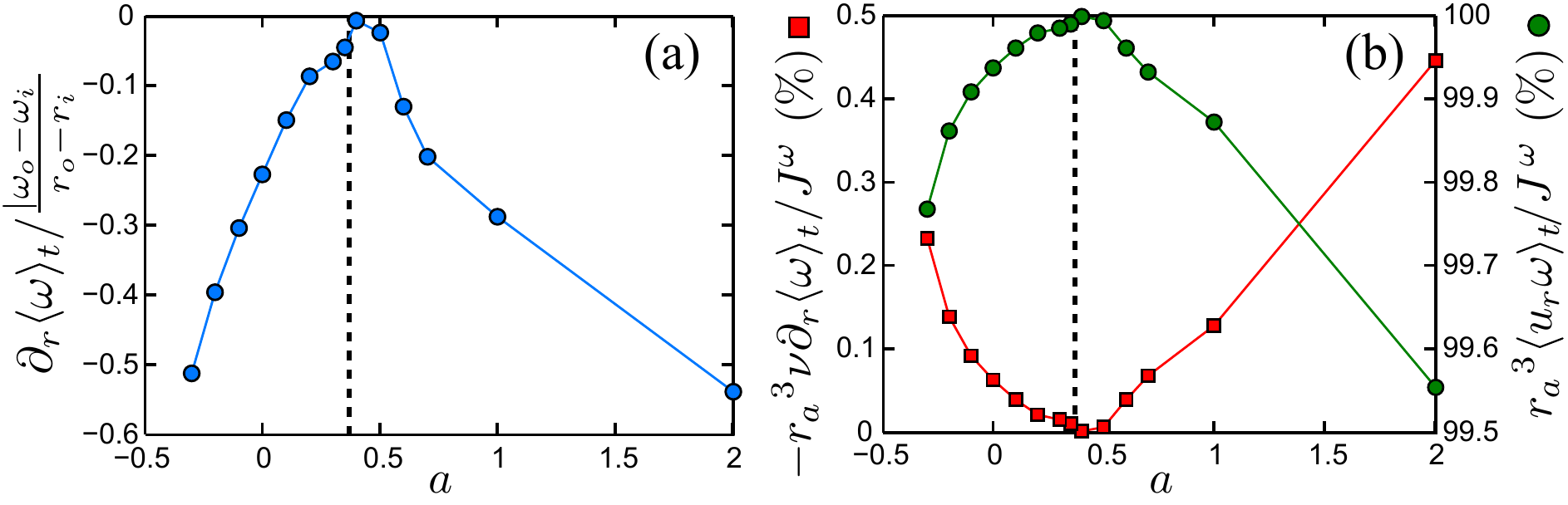}}
\caption{(\emph{a}) Radial gradient $\partial_r \langle \omega \rangle _t$ of the angular velocity profile in the bulk of the flow, non-dimensionalized with the mean $\omega$ slope $\left|\omega_o - \omega_i\right|/(r_o - r_i)$. The values are obtained from figure \ref{fig:12} by fitting a cubic spline to the profiles in order to increase the accuracy of the gradient amplitude estimate. Note that the radial gradients are \emph{negative} throughout, and approach zero when close to $a = a_{bis} \approx \aOpt$. (The connecting lines are guides for the eyes.) (\emph{b}) Resulting ratio of the viscous angular velocity transport term $-r_a^3\nu\partial_r\langle\omega\rangle_{A, t}$ to the total transport $J^\omega$ (squares and left axis) and ratio of the advective angular velocity transport term $r_a^3\langle u_r\omega\rangle_{A, t}$ to the total transport $J^\omega$ (circles and right axis) for the various $a$. These ratios correspond to the second and first terms of (\ref{eq:J_omega}), respectively.
\label{fig:14}}
\end{figure}

From figure \ref{fig:12} it is seen that for nearly all  co- and counter-rotating cases $-0.3 \le a \le 2.0$ the slope of $\left< \omega (\tilde r, t )\right>_t$ is negative. Only around $a= 0.40$ do we find a zero mean angular velocity gradient in the bulk. This case is very close to $\aOpt=0.33\pm0.04$ and $a_{bis}=0.368$. The normalized angular velocity gradient as a function of $a$ is shown in figure \ref{fig:14}(\emph{a}). Indeed, it has a pronounced maximum and zero mean angular velocity gradient very close to $a= a_{bis}\approx \aOpt$, the position of optimal angular velocity transfer. \citet{hou11}, who performed PIV measurements of TC flow in air around $Ta\sim10^{10}$, report a similar trend of a diminishing angular velocity gradient in the center of the TC gap for their investigated $a$ coming from $10.79$ down towards $0.70$, i.e.\ well in the counter-rotating regime. We speculate that their trend would continue and result in a diminishing slope of angular velocity when decreasing $a$ further to their (unreported) case of $\aOpt$. Interestingly, the result of a zero angular velocity gradient across the gap is quite similar to what can be found in Taylor vortex flow, for which the axially averaged circumferential momentum or velocity is nearly uniform across the gap in the bulk (see e.g.\ \cite{mar84, wer94}). We note that in strongly turbulent RB flow the temperature also has a (practically) zero mean gradient in the bulk -- see, for example, the recent review by \cite{ahl09}.

A transition of the flow structure at $a= a_{bis}\approx\aOpt$ can also be confirmed in figure \ref{fig:13}(\emph{a}), in which we have rescaled the mean angular velocity at fixed height as
\be
\langle \tilde{\omega}(\tilde{r}) \rangle_t = \left( \langle \omega (\tilde{r}) \rangle_t  - \omega_o \right) / ( \omega_i - \omega_o ).
\label{eq:omega_norm}
\ee
We observe that up to $a= a_{bis}$ the curves for $\langle \tilde{\omega}(\tilde{r}) \rangle_t$ for all $a$ go through the mid-gap point $(\tilde r = 1/2, \langle \tilde{\omega} (1/2) \rangle_t \approx 0.35)$, implying the mid-gap value $\langle \omega (1/2) \rangle_t \approx 0.35 \omega_i + 0.65 \omega_o$ for the time-averaged angular velocity. However, for $a > a_{bis}$, i.e.\ stronger counter-rotation, the angular velocity at mid-gap becomes larger, as seen in figure \ref{fig:13}(\emph{b}).

Figure \ref{fig:14}(\emph{b}) shows the relative contributions of the molecular and the turbulent transport to the total angular velocity flux $J^{\omega}$ (\ref{eq:J_omega}), i.e.\ for both the diffusive and the advective term. The latter always dominates by far with values beyond 99\%, but at $a \approx  a_{bis}$ the advective term contributes 100\% to the angular velocity flux and the diffusive term nothing, corresponding to the zero mean angular velocity gradient in the bulk at that $a$. This special situation perfectly resembles RB turbulence for which, due to the absence of a mean temperature gradient in the bulk, the whole heat transport is conveyed by the convective term. In the (here unresolved) kinetic boundary layers the contributions just reverse: The convective term strongly decreases if $\tilde{r}$ approaches the cylinder walls at 0 or 1 since $u_r \rightarrow 0$ ($u_z \rightarrow 0$ in RB), while the diffusive term (heat flux in RB) takes over at the same rate, as the total flux $J^{\omega}$ is an $\tilde{r}$-independent constant.

\begin{table}
\begin{center}
  \begin{tabular}{ccccccccccc}
    $a$ & $\psi$ & $Ta$ & $\omega_i$ & $\omega_o$ & $Re_i$ & $Re_o$ & $\mathrm{N_{min}}$ & $\mathrm{N_{max}}$ & $F_{s\mathrm{,min}}$ & $F_{s\mathrm{,max}}$\\
    & (deg.) & $(10^{12})$ & (rad s$^{-1}$) & (rad s$^{-1}$) & $(10^6)$ & $(10^6)$ & $(10^3)$ & $(10^3)$ & (s$^{-1}$) & (s$^{-1}$)\\
    \\
    ~2.000 & ~43.1 & 0.95 & ~16.3 & -32.6~ & 0.26 & -0.74 & 35 & 60 & ~70 & 2.4$k$\\
    ~1.000 & ~27.2 & 1.06 & ~25.8 & -25.8~ & 0.42 & -0.58 & 26 & 60 & ~52 & 1.6$k$\\
    ~0.700 & ~17.2 & 1.12 & ~31.2 & -21.9~ & 0.51 & -0.49 & 55 & 80 & ~46 & 0.7$k$\\
    ~0.600 & ~12.8 & 1.15 & ~33.6 & -20.1~ & 0.54 & -0.46 & 67 & 80 & ~56 & 0.5$k$\\
    ~0.500 & ~~7.7 & 1.20 & ~36.4 & -18.2~ & 0.59 & -0.41 & 31 & 60 & ~63 & 0.6$k$\\
    ~0.400 & ~~2.0 & 1.22 & ~39.5 & -15.8~ & 0.64 & -0.36 & 10 & 25 & ~57 & 0.7$k$\\
    ~0.350 & ~-1.2 & 1.25 & ~41.4 & -14.5~ & 0.67 & -0.33 & 11 & 25 & ~45 & 0.5$k$\\
    ~0.300 & ~-4.5 & 1.28 & ~43.6 & -13.1~ & 0.70 & -0.30 & 12 & 25 & ~64 & 0.8$k$\\
    ~0.200 & -11.6 & 1.34 & ~48.3 & ~-9.65 & 0.78 & -0.22 & 14 & 25 & ~78 & 1.0$k$\\
    ~0.100 & -19.2 & 1.42 & ~54.2 & ~-5.46 & 0.88 & -0.12 & 17 & 25 & ~93 & 1.3$k$\\
    ~0.000 & -27.2 & 1.53 & ~61.8 & ~~0.00 & 1.00 & ~0.00 & 13 & 25 & ~72 & 2.3$k$\\
    -0.100 & -35.2 & 1.67 & ~71.9 & ~~7.21 & 1.16 & ~0.16 & 25 & 25 & 147 & 1.7$k$\\
    -0.200 & -42.8 & 1.87 & ~85.7 & ~17.1~ & 1.39 & ~0.39 & 22 & 25 & 121 & 1.6$k$\\
    -0.300 & -50.0 & 2.21 & 106.0 & ~31.9~ & 1.72 & ~0.72 & ~6 & 25 & ~32 & 7.5$k$\\
  \end{tabular}
  \caption{LDA experimental conditions. Each case of $a$ is examined at fixed $Re_i - Re_o = 1.0\times10^6$. Here $a, \psi, Ta, \omega_{i,o}, Re_{i,o}$ are as defined before; see also table \ref{tab:G-data}. The minimum and maximum number of detected LDA bursts along the radial scan is given by $\mathrm{N_{min}}$ and $\mathrm{N_{max}}$, respectively. Likewise for the minimum and maximum average burst rate $F_{s\mathrm{,min}}$ and $F_{s\mathrm{,max}}$.}
  \label{tab:LDA_experiments}
  \end{center}
\end{table}

\begin{figure}
\centerline{\includegraphics[width=8cm]{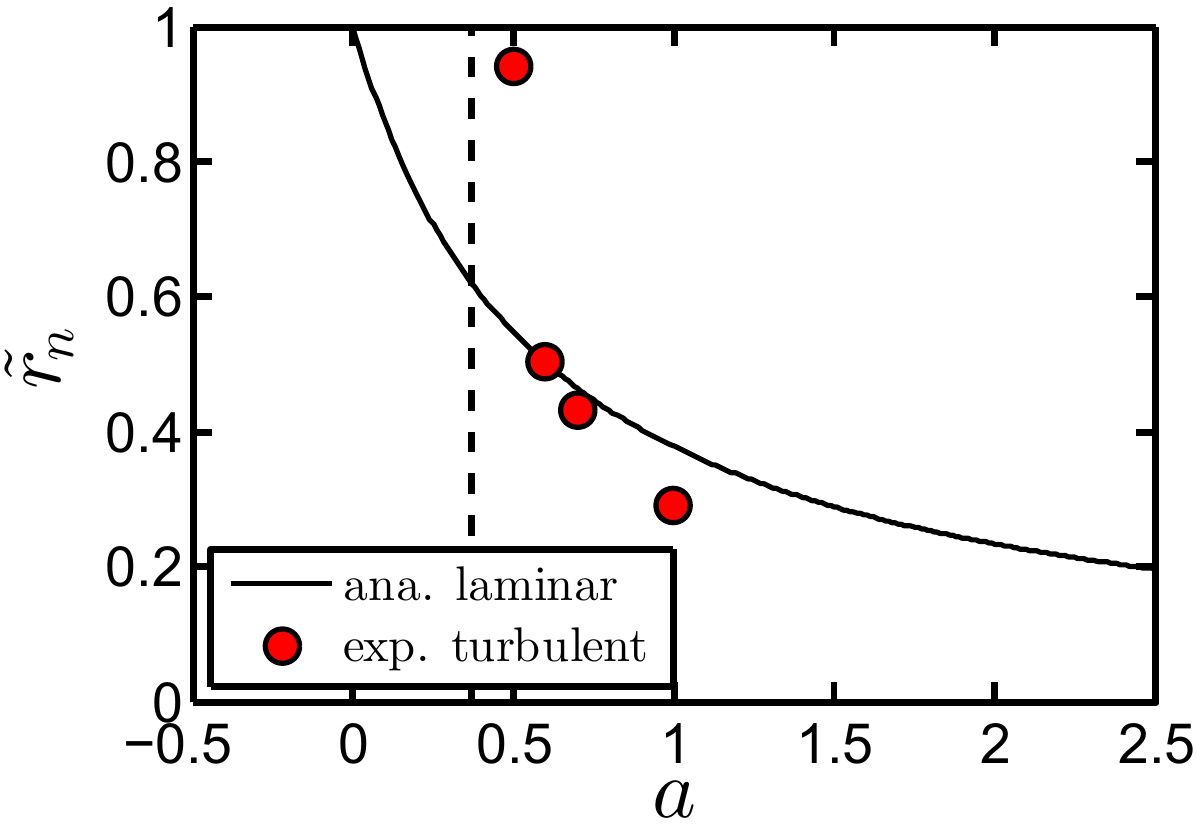}}
\caption{The measured radial position $\tilde{r}_n = (r_n - r_i) / (r_o - r_i)$ of the neutral line defined by $\langle \omega (\tilde r_n) \rangle_t = 0$ at fixed height $z/L = 0.44$ as a function of $a$, indicated by the circles. The values are obtained from figure \ref{fig:12} by fitting a cubic spline to the profiles in order to increase the accuracy of locating $\tilde{r}_n$, resulting in an accuracy equal to the symbol size. The straight vertical line corresponds to $a_{bis} = 0.368$, below which the neutral line is within the outer BL and cannot be resolved by our LDA technique. The solid line corresponds to the neutral line in the laminar case calculated analytically with (\ref{eq:laminar_profile}). Note that for significant counter-rotation $a=1$ and for this particular height, the neutral line in the turbulent case lies farther inside, nearer to the inner cylinder, than in the laminar flow case.
\label{fig:15}}
\end{figure}

From the measurements presented in figure \ref{fig:12}, we can extract the neutral line $\tilde r_n$, defined by $\left< \omega (\tilde r_n ,t ) \right>_t = 0$ at fixed height $z/L=0.44$ for the turbulent case. The neutral line is, of course, part of a neutral \emph{surface} throughout the TC volume. We expect that, for large $a\gg \aOpt$, the flow will be so much stabilized that an axial dependence of the location of the neutral line shows up, in spite of the large $Ta$ numbers, in contrast to the cases $a \approx \aOpt$ where we expect the location of the neutral line to be axially independent for large enough $Ta$. The results for $\tilde r_n$ are shown in figure \ref{fig:15}. Note that, while for $a\le 0$ (co-rotation) there obviously is no neutral line at which $\langle \omega \rangle_t = 0$, for $a > 0$ a neutral line exists at some position $\tilde r_n > 0$. As long as it is still within the outer kinematic BL, we cannot resolve it. This turns out to be the case for $0 < a \lesssim {a_{bis}}$. But for ${a_{bis}} \lesssim a$ the neutral line can be observed within the bulk and is well resolved with our measurements. So, again, we see two regimes: for $0 < a \lesssim a_{bis}$ in the laminar case the stabilizing outer cylinder rotation shifts the neutral line inwards, but, due to the now {\it free boundary} between the stable outer $r$ range and the unstable range between the neutral line and the inner cylinder, the flow structures extend beyond $\tilde r_n$. Thus also in the turbulent flow case the unstable range flow extends to the close vicinity of the outer cylinder. The increased shear and the strong turbulence activity originating from the inner cylinder rotation are too strong and prevent the neutral line being shifted off the outer kinematic BL. As described in \cite{ess96}, this is the very mechanism that shifts the minimum of the viscous instability curve to the left of the $Re_i =0$ axis. Therefore, the observed behaviour of the neutral line position as a function of $a$ is another confirmation of the above idea that $\aOpt$ coincides with the angle bisector $a_{bis}$ of the instability range in parameter space. The small-$Re_o$ and the large-$Re_o$ behaviours perfectly merge. Only for $a> \aOpt$ is the stabilizing effect from the outer cylinder rotation strong enough, and the width of the stabilized range broad enough, so that a neutral line $\tilde{r}_n$ can be detected in the bulk of the TC flow. This behaviour is similar to what is reported by \citet{hou11}, see their figure 6.

One would expect that for much weaker turbulence $Ta \ll 10^{11}$, the capacity of the turbulence around the inner cylinder to push the neutral line outwards would decrease, leading to a smaller $a_{opt}$ for these smaller Taylor numbers. Numerical simulations by H.\ Brauckmann and B.\ Eckhardt of the University of Marburg ($Ta$ up to $1 \times 10^9$) and independently ongoing DNS by \cite{ost12} of the University of Twente (presently $Ta$ up to $1 \times 10^8$) seem to confirm this view.

For much weaker turbulence, one would also expect a more pronounced height dependence of the neutral line, which will be pushed outwards where the Taylor rolls are going outwards and inwards where they are going inwards. Based on our height dependence studies of section \ref{sec2}, we expect  that this height dependence will be much weaker or even fully washed out in the strongly turbulent regime $Ta > 10^{11}$ in which we operate the TC apparatus. However, in figure \ref{fig:15} we observe that the neutral line in the turbulent case lies more inside than in the laminar case, and this result is difficult to rationalize apart from assuming some axial dependence of the neutral line location, i.e.\ more outwards locations of the neutral line at larger and smaller height. In future work we will study the axial dependence of the neutral line in the turbulent and counter-rotating case in more detail.

\begin{figure}
\centerline{\includegraphics[width=.75\textwidth]{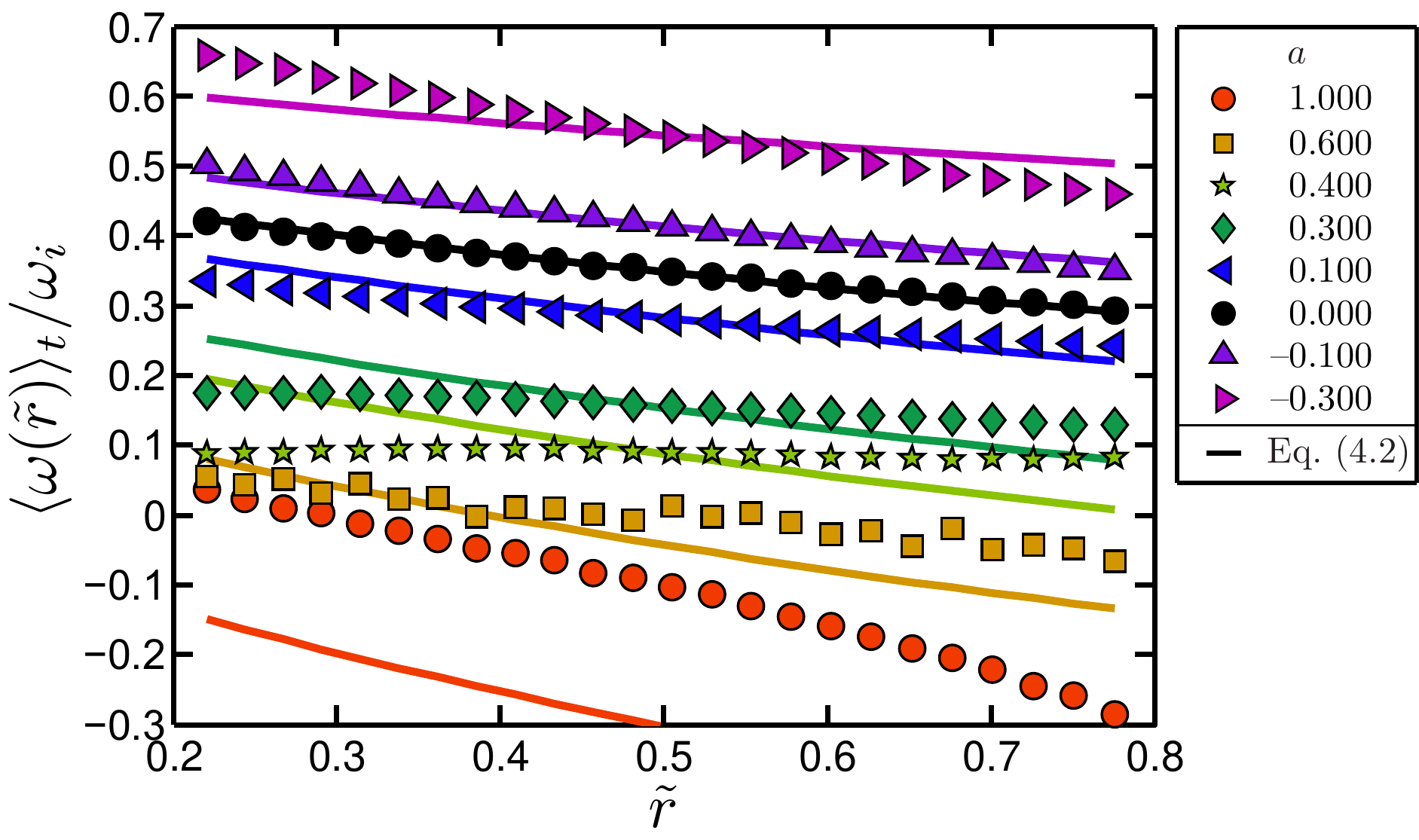}}
\caption{Normalized angular velocity  $\langle\omega(\tilde{r})\rangle_t/\omega_i$ versus normalized gap distance $\tilde{r}$ for various $a$. The symbols indicate the experimental data as already presented in figures \ref{fig:12} and \ref{fig:13}(\emph{a}). To improve visual appearance, only a selection of the $a$ cases is shown. The thick solid lines are given by (\ref{eq:Busse}) for $\eta=0.716$.
\label{fig:16}}
\end{figure}

For completeness, we also compare our experimental angular velocity profiles with those employed in the upper-bound theory by \citet{bus72}. \citet{bus72} derived an expression for the angular velocity profiles in the limit of infinite Reynolds number. Translating that expression to the notation used in the present work gives
\be
\omega_\mathrm{Busse} = \frac{\omega_i - \omega_o}{r^2}\left[\frac{{r_i}^2}{4(1-\eta^2)}\right] + \omega_i\left[\frac{\eta^2-2\eta^4}{2-2\eta^4} \right] + \omega_o\left[\frac{2-\eta^2}{2-2\eta^4}\right].
\label{eq:Busse}
\ee
As already shown by \citet{lew99}, for the case $a=0$ excellent agreement between the Busse profile and the experimental one is found. However, for $a$ farther away from zero, there is a greater discrepancy between the experimental data and the profiles suggested by the upper-bound theory, as shown in figure \ref{fig:16}(\emph{a}). When we rescale the angular velocity profiles to $\tilde{\omega}$, according to (\ref{eq:omega_norm}), the profiles as given by the upper-bound theory \citep{bus72} fall on top of each other for all $a$,
\be
\tilde{\omega}_\mathrm{Busse}=\frac{\omega_\mathrm{Busse} - \omega_o}{\omega_i - \omega_o}=\frac{{r_i}^2}{4\left({r_o}^4-{r_i}^4\right)}\left[\frac{{r_o}^2\left({r_i}^2+{r_o}^2\right)}{r^2} +2{r_o}^2-4{r_i}^2 \right]
\label{eq:Busse2}
\ee
In contrast to the collapsing upper-bound profiles, the experimental data in figure \ref{fig:16}(\emph{b}) show a different trend. Clearly, the profiles suggested by the upper bound theory are in general not a good description of the physically realized profiles, apart from the $a=0$ case. Given the complexity of the flow, this may not be surprising.

While in this section we have only focused on the time-mean values of the angular velocity, in the following section we will give more details on the probability density functions (p.d.f.) in the two different regimes below and above $a_{bis}$ and thus on the different dynamics of the flow in these two different
regimes.


\section{Turbulent flow organization in the gap between the cylinders}\label{sec5}

\begin{figure}
\centerline{\includegraphics[width=1\textwidth]{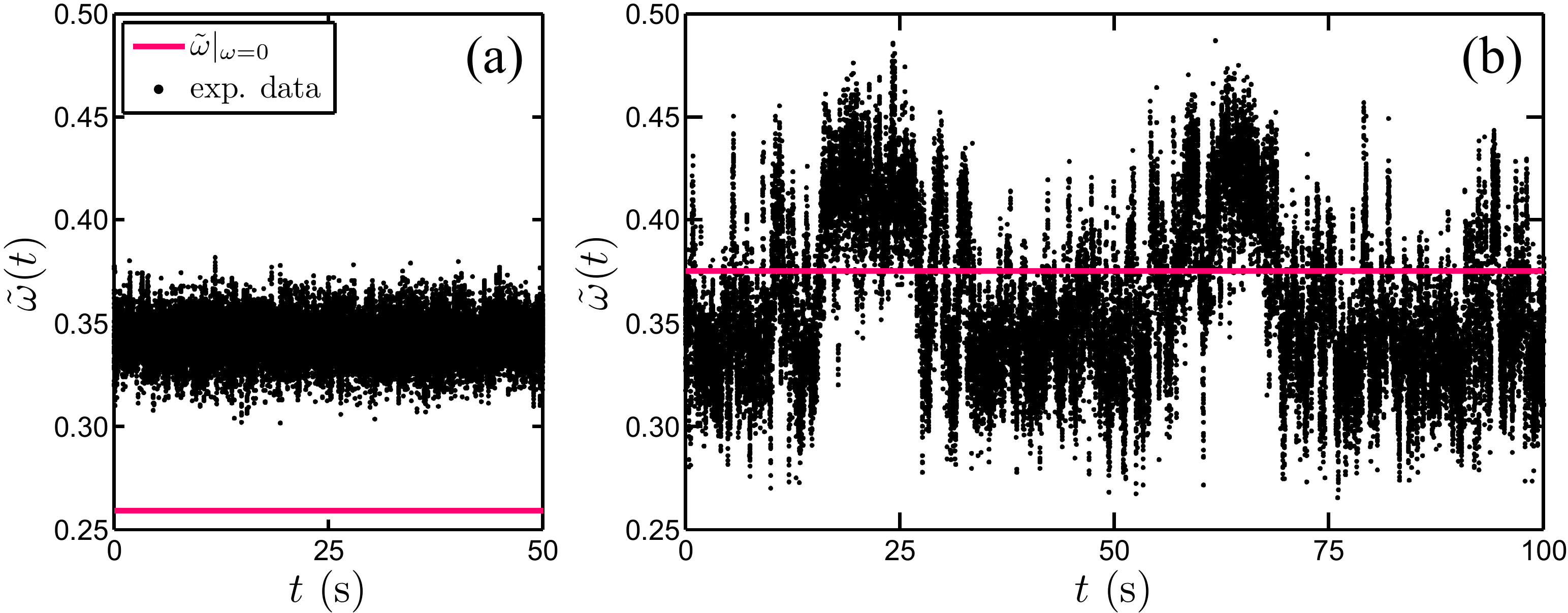}}
\caption{Time series of the dimensionless angular velocity $\tilde{\omega}(t)$ as defined in (\ref{eq:omega_norm}) but without $t$-averaging, acquired with LDA. (\emph{a}) For $a = 0.35$, co-rotation dominates; at $\tilde{r}=0.60$ with an average data acquisition rate of 456 Hz. (\emph{b}) For $a = 0.60$, counter-rotation dominates; same $\tilde{r}=0.60$ with an average data acquisition rate of 312 Hz. Panel (\emph{a}) shows a unimodal velocity distribution whereas panel (\emph{b}) reveals a bimodal distribution interpreted as being caused by intermittent bursts out of the unstable inner regime with angular velocity between $\omega_i$ and $\omega = 0$, i.e.\ $(\omega_o - 0)/(\omega_o - \omega_i) < \tilde{\omega} < 1$, into the stable outer regime with angular velocity between $\omega = 0$ and $\omega_0$, i.e.\ $0 < \tilde{\omega} < (\omega_o - 0)/(\omega_o - \omega_i)$. The solid grey (red-pink) line indicates the neutral line $\omega = 0$, corresponding to {$\tilde \omega = \omega_o/ (\omega_o-\omega_i) = a / (1+a)$}, for the specific $a$.
\label{fig:17}}
\end{figure}

\begin{figure}
\centerline{\includegraphics[width=1\textwidth]{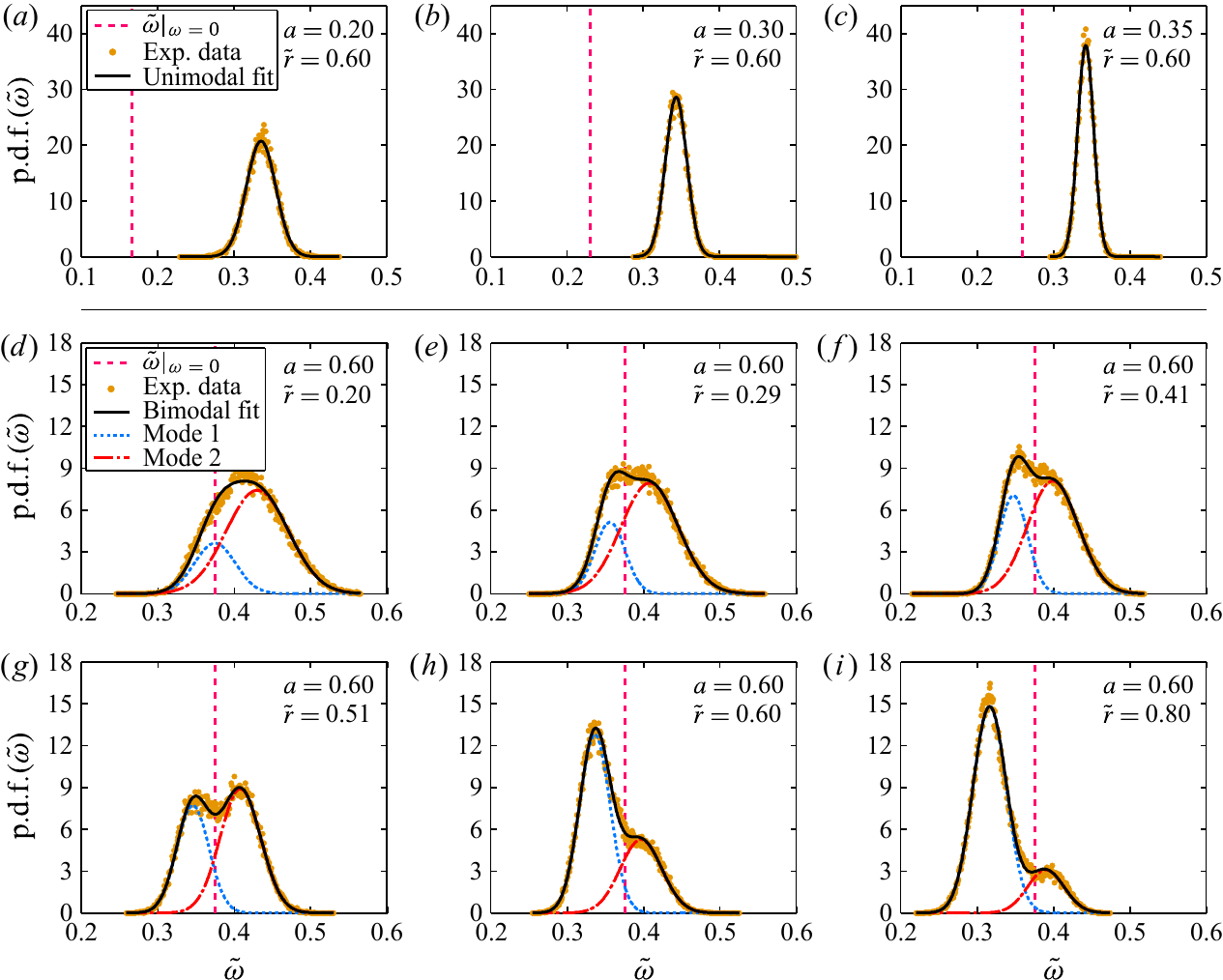}}
\caption{Probability density functions of the angular velocity $\tilde{\omega}(t)$ distributions for various cases of $a$ and position $\tilde{r} = (r - r_i) / (r_o - r_i)$:
(\emph{a}) $a=0.20$, $\tilde r = 0.60$;
(\emph{b}) $a=0.30$, $\tilde r = 0.60$; and
(\emph{c}) $a=0.35$, $\tilde r = 0.60$.
The other panels all show $a=0.60$ but at different non-dimensional gap distances $\tilde{r}$:
(\emph{d}) $0.20$;
(\emph{e}) $0.29$;
(\emph{f}) $0.41$;
(\emph{g}) $0.51$;
(\emph{h}) $0.60$; and
(\emph{i}) $0.80$.
The gold circles indicate the measured distribution obtained by LDA. While for panels (\emph{a--c}) one Gaussian distribution (black solid curve) describes the data well, for panels (\emph{e--i}) a superposition of two Gaussians is needed for a good fit (dotted blue and dashed-dotted red curves).  We call the two Gaussians the two `modes' of the flow. The fitting algorithm gives the mean, the standard deviation and the mixture coefficient of modes 1 and 2, which recombine to the black solid line, describing the measured distribution well. The dashed red-pink vertical line shows the neutral line $\omega = 0$, implying {$\tilde \omega = \omega_o/ (\omega_o-\omega_i) = a/(1+a)$}.
\label{fig:18}}
\end{figure}

Time series of the angular velocity at $\tilde r= 0.60$ below the optimum amplitude at $a = 0.35 < a_{bis} = 0.368$ (co-rotation dominates) and above the optimum at $a = 0.60 > a_{bis} = 0.368$ (counter-rotation dominates) are shown in figure \ref{fig:17}(\emph{a, b}), respectively. While in the former case we always have $\omega (t)  > 0$ and a Gaussian distribution (see figure \ref{fig:18}\emph{a}), in the latter case we find a bimodal distribution with one mode fluctuating around a positive angular velocity  and one mode fluctuating around a negative angular velocity. This bimodal distribution of $\omega (t)$ is confirmed in various p.d.f.s shown in figure \ref{fig:18}. We interpret this intermittent behaviour of the time series as an indication of turbulent bursts originating from the turbulent region in the vicinity of the inner cylinder and penetrating into the stabilized region near the outer cylinder. We find such bimodal behaviour for all $a >  a_{bis}$ (see figure \ref{fig:18}\emph{d--i}), whereas for $a< a_{bis}$ we find a unimodal behaviour (see figure \ref{fig:18}\emph{a--c}). Apart from one case ($a = 0.50$) we do not find any long-time periodicity of the bursts in $\omega (t)$. In future work we will perform a full spectral analysis of long time series of $\omega(t)$ for various $a$ and $\tilde r$.

\begin{figure}
\centerline{\includegraphics[width=1\textwidth]{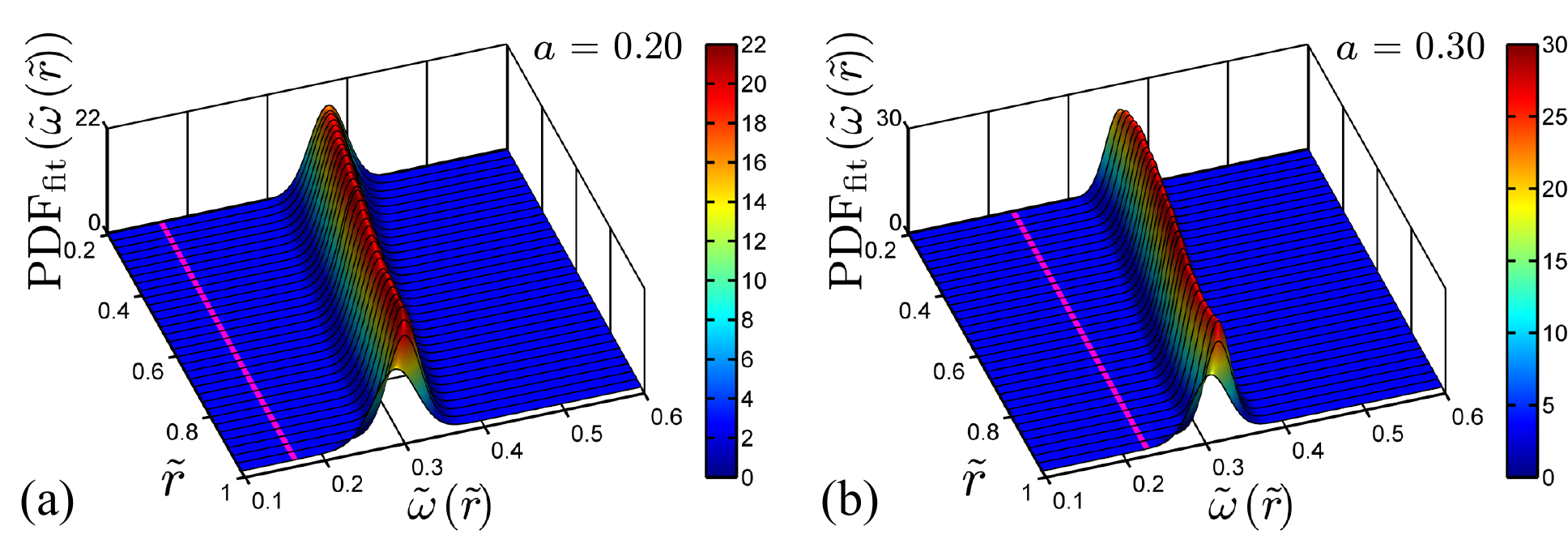}}
\caption{Three-dimensional visualization of the (normalized) angular velocity p.d.f. with a continuous scan of $\tilde r$ for two unimodal cases, (\emph{a}) $a=0.20$ and (\emph{b}) $a=0.30$, both being smaller than $a_{bis} = 0.368$. The red-pink line corresponds to the neutral line $\omega = 0$, i.e.\ \ $\tilde \omega = \omega_o/ (\omega_o-\omega_i) = a / (1+a)$.
\label{fig:19}}
\end{figure}

\begin{figure}
\centerline{\includegraphics[width=1\textwidth]{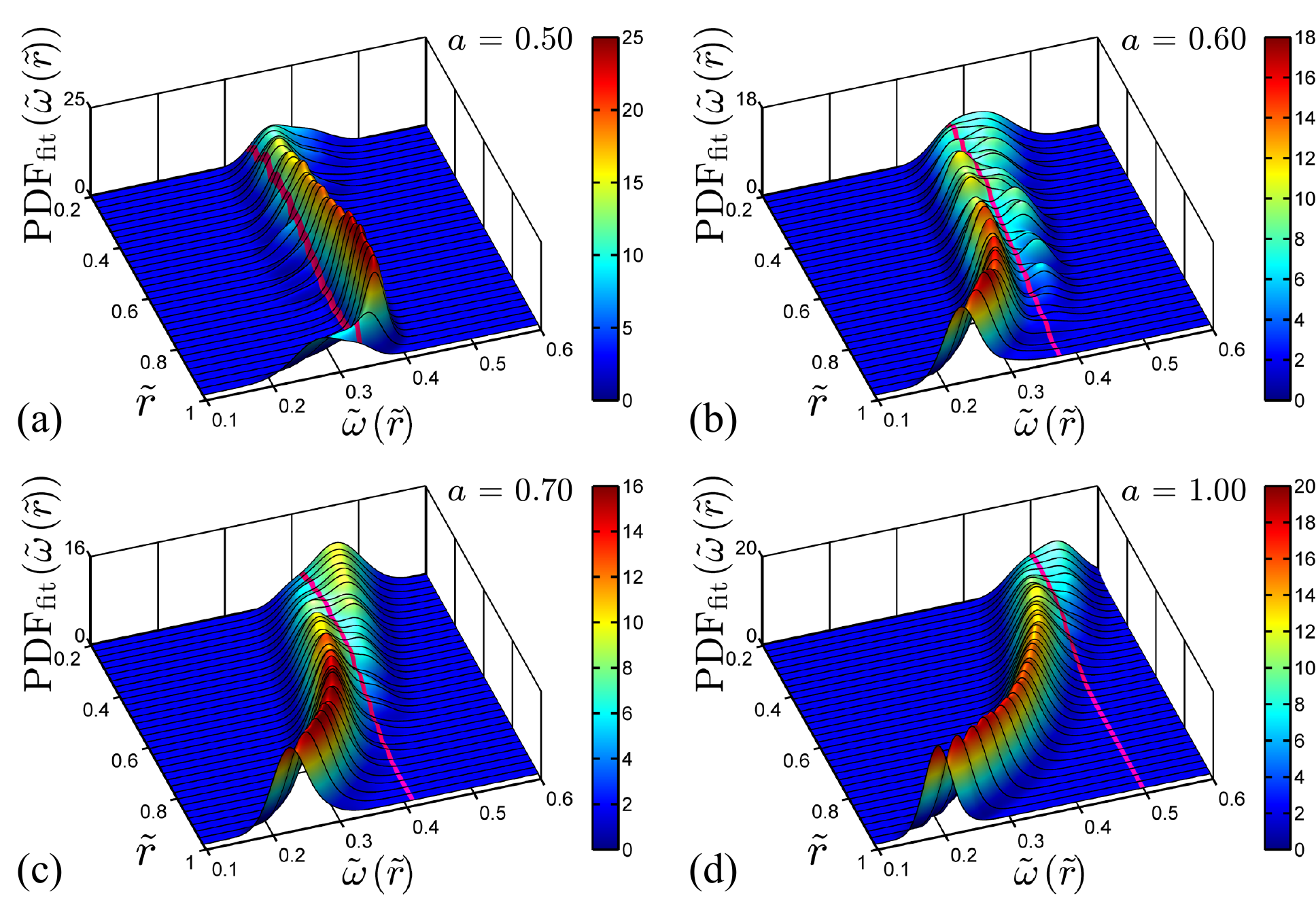}}
\caption{Same as in figure \ref{fig:19}, but now for the bimodal cases $a> a_{bis} = 0.368$:
(\emph{a}) $a=0.50$;
(\emph{b}) $a=0.60$;
(\emph{c}) $a=0.70$; and
(\emph{d}) $a=1.00$.
The bimodal character with one mode being left of the neutral line $\omega = 0$ and the other mode being right of the neutral line becomes particularly clear for panels (\emph{b}) and (\emph{c}).
\label{fig:20}}
\end{figure}

Three-dimensional visualizations of the $\omega (t)$ p.d.f. for all $0< \tilde r< 1$ are provided in figure \ref{fig:19} for unimodal cases $a< a_{bis}$ and in figure \ref{fig:20} for bimodal cases $a> a_{bis}$. In the latter figure the switching of the system between positive and negative angular velocity becomes visible.

\begin{figure}
\begin{minipage}[t]{0.5\linewidth}
\centering
\includegraphics[width=1\textwidth]{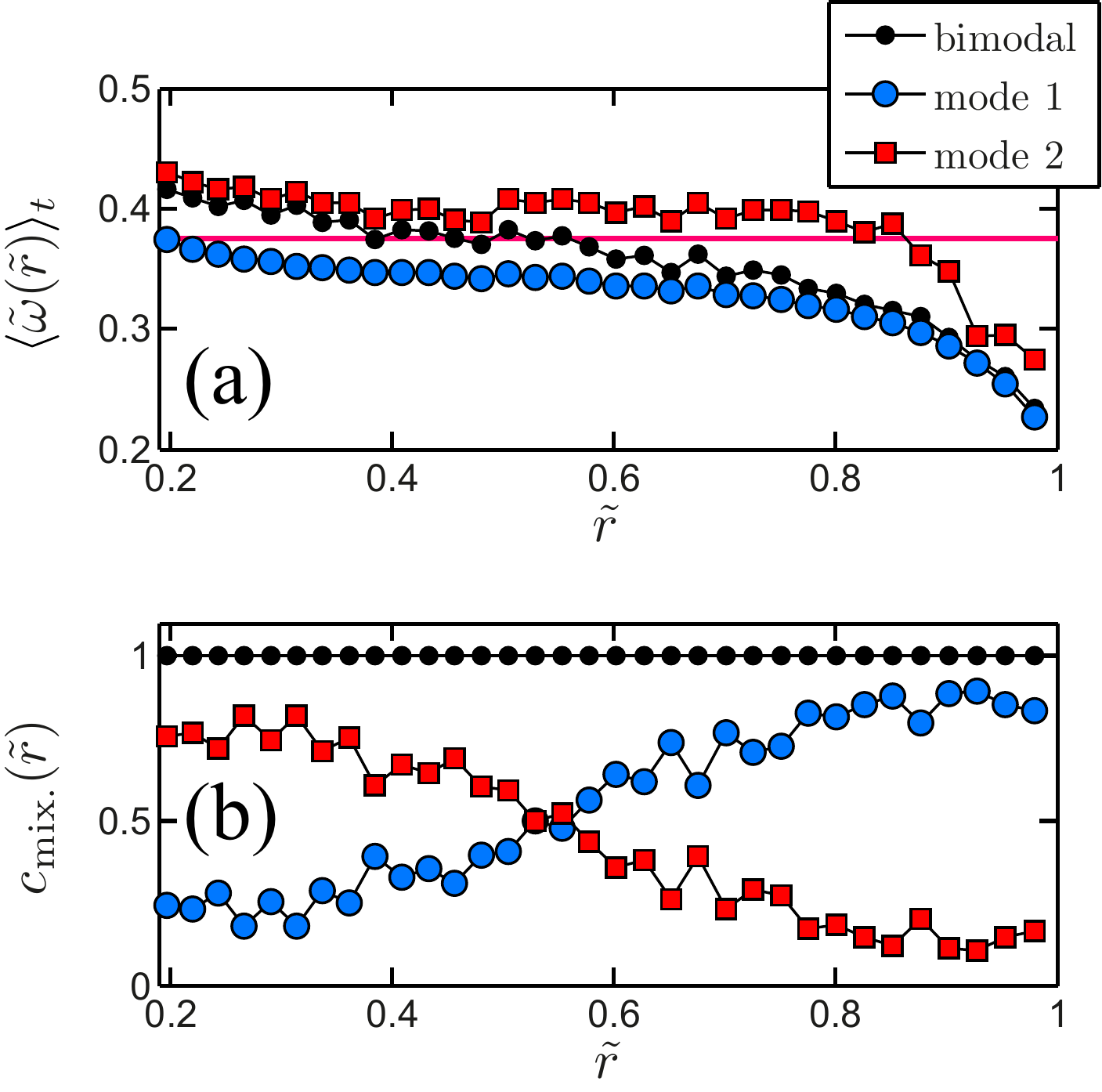}
\caption{(\emph{a}) The time-averaged, normalized angular velocity $\left< \tilde \omega (\tilde r ,t )\right>_t$ (black circles) and that of the two individual modes (blue circles and red squares) for $a = 0.60$. (\emph{b}) The mixing coefficient $c_\mathrm{mix.}(\tilde{r})$ defined as the relative contribution of the area underneath the p.d.f. of the respective individual mode with respect to the total p.d.f. area.
\label{fig:21}}
\end{minipage}
\hspace{0.2cm}
\begin{minipage}[t]{0.5\linewidth}
\centering
\includegraphics[width=1\textwidth]{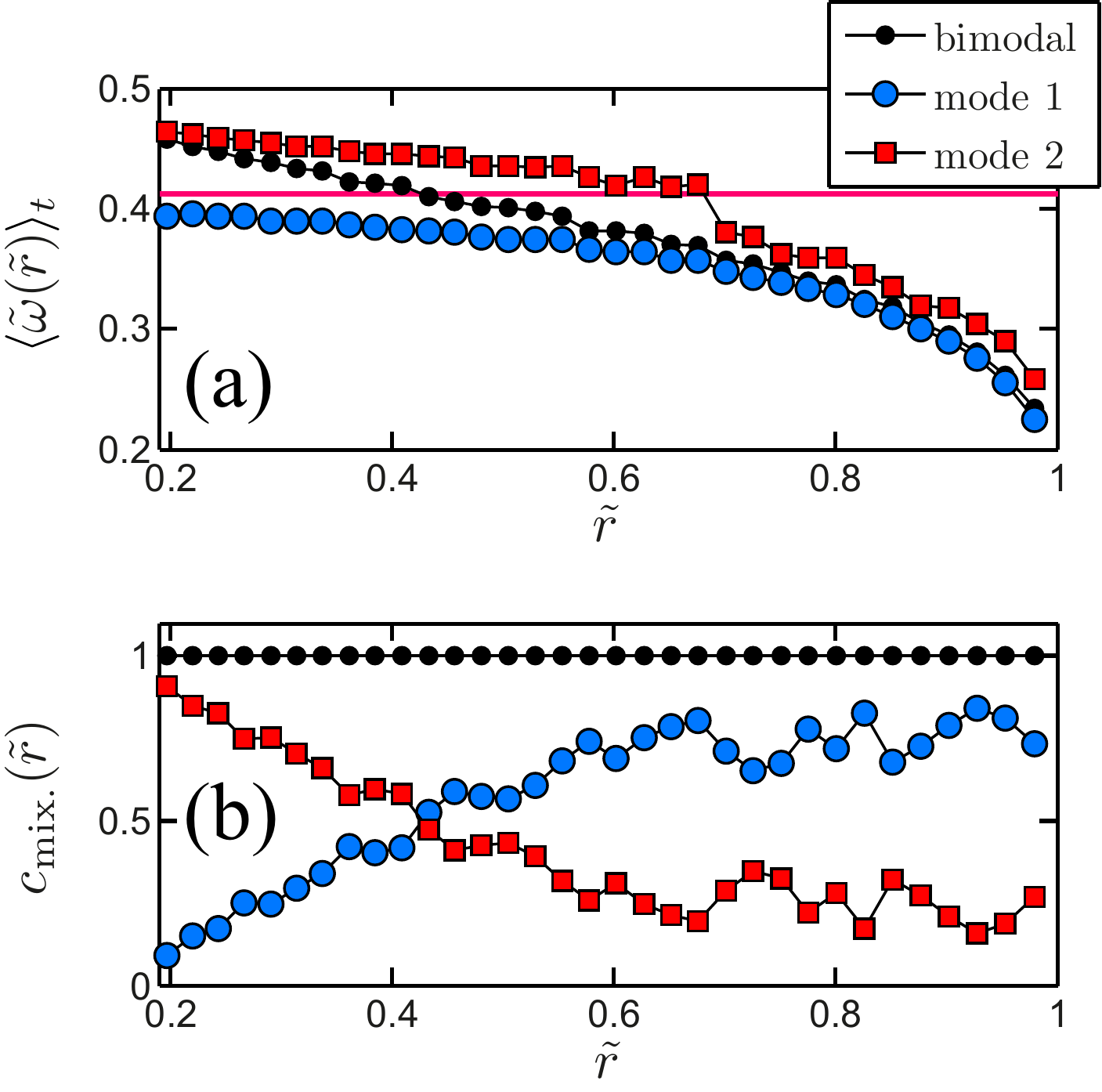}
\caption{Same as in figure \ref{fig:21}, but now for $a = 0.70$.
\label{fig:22}}
\end{minipage}
\end{figure}

Further details on the two observed individual modes, such as their mean and their mixing coefficient, are given in figure \ref{fig:21} (for $a= 0.60$) and figure \ref{fig:22} (for $a=0.70$), both well beyond $a_{bis}$. In both cases one observes that the contribution from the large-$\omega$ mode  (dashed-dotted red curve, $\omega >0$, apart from positions close to the outer cylinder) is, as expected, highest at the inner cylinder and fades away when going outwards, whereas the small-$\omega$ mode (dotted blue curve) has the reverse trend. Note, however, that, even at $\tilde r=0.20$, e.g.\ relatively close to the inner cylinder, there are moments for which $\omega$ is negative, i.e.\ patches of stabilized liquids are advected inwards, just as patches of turbulent flow are advected outwards.

This mechanism resembles the angular velocity exchange mechanism suggested by \citet{cou96} just beyond the onset of turbulence. These authors suggest that for the counter-rotating case there is an outer region that is centrifugally stable, but subcritically unstable, thus vulnerable to distortions coming from the centrifugally unstable inner region. The inner and outer regions are separated by the neutral line. For the low $Re$ of \citet{cou96} the inner region is not yet turbulent, but displays interpenetrating spirals, i.e.\ a chaotic flow with various spiral Taylor vortices. For our much larger Reynolds numbers, the inner flow will be turbulent and the distortions propagating into the subcritically unstable outer regime will be turbulent bursts. These then will lead to intermittent instabilities in the outer regime. In future work, these speculations must further be quantified.

\section{Summary, discussion, and outlook}\label{sec7}

In conclusion, we have experimentally explored strongly turbulent TC flow with $Ta > 10^{11}$ in the co- and counter-rotating regimes. We find that, in this large-Taylor-number $Ta$ regime and well off the instability lines, the dimensionless angular velocity transport flux within experimental precision can be written as $Nu_\omega (Ta, a) = f(a) Ta^\gamma$ with (within our accuracy) a universal $\gamma = 0.39 \pm 0.03$ for all $a$. This is the effective scaling exponent of the ultimate regime of TC turbulence predicted by \cite{gro11} for RB flow and transferred to TC by the close correspondence between RB and TC elaborated in the EGL theory. When starting off counter-rotation, i.e.\ when increasing $a$ beyond zero, the angular velocity flux does not reduce but instead is first further enhanced, due to the enhanced shear, before finally, beyond $a = \aOpt \approx  0.33$, the stabilizing effect of the counter-rotation leads to a reduction of the angular velocity transport flux $Nu_{\omega}$. Around $a_{opt}$  the mean angular velocity profile was shown to have zero gradient in the bulk for the present large $Ta$. Despite already significant counter-rotation for $0 < a \lesssim \aOpt$, there is no neutral line outside the outer BL; furthermore, the probability distribution function of the angular velocity has only one mode. For larger $a$, beyond $a \gtrsim \aOpt$, a neutral line can be detected in the bulk and the p.d.f. here becomes bimodal, reflecting intermittent bursts of turbulent patches from the turbulent inner $r$ regime towards the stabilized outer $r$ regime. We offered a hypothesis that gives a unifying view and consistent understanding of all these various findings.

Clearly, the present study is only the start of a long experimental program to further explore turbulent TC flow. Much more work still must be done. In particular, we mention the following open issues:
\begin{itemize}
\item
For strong counter-rotation, i.e.\ large $a$, the axial dependence must be studied in much more detail. This holds in particular for the location of the neutral line. We expect that, for large $a \gg \aOpt$, the flow will be stabilized so much that an axial dependence of the location of the neutral line shows up again, in spite of the large $Ta$ numbers. It is then more appropriate to speak of a neutral \emph{surface}.
\item
Modern PIV techniques should enable us to directly resolve the inner and outer BLs.
\item
With the help of PIV studies we also hope to visualize BL instabilities and thus better understand the dynamics of the flow, in particular, in the strongly counter-rotating case. The key question is: How does the flow manage to transport angular velocity from the turbulent inner regime towards the outer cylinder -- thereby crossing the stabilized outer regime?
\item
The studies must be repeated at lower $Ta$ to better understand the transition from weakly turbulent TC towards the ultimate regime as apparently seen in the present work. In our present set-up we can achieve such a regime with silicone oil instead of water.
\item
One must also better understand the $a$ dependence of the transition to the ultimate regime.
\item
All these studies should be complemented with direct numerical simulation of co- and counter-rotating TC flow. Just as has happened in recent years in turbulent RB flow (cf.\ \cite{ste10,ste11}), also for turbulent TC flow we expect to narrow the gap between numerical simulation and experiment, or even to close it for not too turbulent cases, allowing for one-to-one comparisons.
\item
Obviously, our studies must be extended to different $\eta$ values, in order to check the hypothesis (\ref{eq:a_bis}) and to see how this parameter affects the flow organization.
\end{itemize}
Clearly, many exciting discoveries and wonderful work with turbulent TC flow is ahead of us.

\begin{acknowledgements}
This study was financially supported by the Technology Foundation STW of the Netherlands, which is financially supported by NWO. The authors would like to thank G.\ Ahlers, H.\ Brauckmann, F.H.\ Busse, B.\ Eckhardt, D.P.\ Lathrop, R. Ostilla M\'onico, M.S.\ Paoletti and G. Pfister for scientific discussions. D. Lohse and S.\ Grossmann also thank F.H.\ Busse for pointing them to \cite{bus72} of which they were not aware when writing \cite{eck07b}. We also gratefully acknowledge technical contributions from TCO-TNW Twente, G.-W.\ Bruggert, M.\ Bos, and B.\ Benschop.
\end{acknowledgements}

\bibliographystyle{jfm}

\end{document}